\documentclass[journal,comsoc]{IEEEtran}

\usepackage[T1]{fontenc}
\usepackage{dblfloatfix}
\usepackage{cite}
\usepackage{amsmath,amssymb,amsfonts}
\usepackage{algorithmic}
\usepackage{graphicx}
\usepackage{mathtools}
\usepackage{textcomp}
\usepackage{multirow}   
\usepackage{makecell}
\usepackage{array}
\usepackage{amsmath}
\usepackage{caption}
\usepackage[table]{xcolor}
\usepackage{xcolor}
\newcommand{\dd}[1]{\mathrm{d}#1}
\def\BibTeX{{\rm B\kern-.05em{\sc i\kern-.025em b}\kern-.08em
    T\kern-.1667em\lower.7ex\hbox{E}\kern-.125emX}}
    
\ifCLASSINFOpdf

\else

\fi

\usepackage[cmintegrals]{newtxmath}

\begin{document}
%
\title{An Open Source Modeling Framework for Interdependent Energy-Transportation- Communication Infrastructure in Smart and Connected Communities}

\author{Xing Lu\textsuperscript{1}, Kathryn Hinkelman\textsuperscript{1}, Yangyang Fu\textsuperscript{1}, Jing Wang\textsuperscript{1}, Wangda Zuo\textsuperscript{1}, Qianqian Zhang\textsuperscript{2}, Walid Saad\textsuperscript{2}
\thanks{\textsuperscript{1} Xing Lu, Kathryn Hinkelman, Yangyang Fu, Jing Wang, Wangda Zuo were with Department of Civil, Environmental and Architectural Engineering, University of Colorado, Boulder, CO 80309 USA, e-mail:   \textbraceleft Xing.Lu-1, Kathryn.Hinkelman, Yangyang.Fu, Jing.Wang, Wangda.Zuo\textbraceright @colorado.edu} 
\thanks{\textsuperscript{2} Qianqian Zhang, Walid Saad were with Electrical and Computer Engineering Department, Virginia Tech, Blacksburg, VA 24061 USA, e-mail: \textbraceleft qqz93,walids\textbraceright @vt.edu.}
\thanks{This research was supported by the National Science Foundation under Awards No. IIS-1802017 and IIS-1633363. BIGDATA: Collaborative Research: IA: Big Data Analytics for Optimized Planning of Smart, Sustainable, and Connected Communities}}

%
%

\markboth{April 7~2019}%
{Lu \MakeLowercase{\textit{et al.}}: An Open Source Modeling Framework for Interdependent Energy-Transportation- Communication Infrastructure in Smart and Connected Communities}
%



\maketitle

\begin{abstract}
Infrastructure in future smart and connected communities is envisioned as an aggregate of public services, including the energy, transportation and communication systems, all intertwined with each other.
The intrinsic interdependency among these systems may exert underlying influence on both design and operation of the heterogeneous infrastructures.
However, few prior studies have tapped into the interdependency among the three systems in order to quantify their potential impacts during standard operation.
In response to this, this paper proposes an open source, flexible, integrated modeling framework suitable for designing coupled energy, transportation, and communication systems and for assessing the impact of their interdependencies.
First, a novel multi-level, multi-layer, multi-agent approach is proposed to enable flexible modeling of the interconnected energy, transportation, and communication systems.
Then, for the framework's proof-of-concept, preliminary component and system-level models for different systems are designed and implemented using Modelica, an equation-based object-oriented modeling language.
Finally, three case studies of gradually increasing complexity are presented (energy, energy + transportation, energy + transportation + communication) to evaluate the interdependencies among the three systems.
Quantitative analyses show that the deviation of the average velocity on the road can be 10.5\% and the deviation of the power draw from the grid can be 7\% with or without considering the transportation and communication system at the peak commute time, indicating the presence of notable interdependencies.
The proposed modeling framework also has the potential to be further extended for various modeling purposes and use cases, such as dynamic modeling and optimization, resilience analysis, and integrated decision making in future connected communities.
\end{abstract}


%
\IEEEpeerreviewmaketitle

\section{Introduction}
\label{sec:introduction}
\IEEEPARstart{U}{rbanization} has become a mega-trend in the world today \cite{b1}.
The resulting large population in urban communities will exert tremendous pressure on existing infrastructure.
To mitigate this issue, the concept of smart and connected communities has recently been proposed in which new and green technologies are embraced collectively to deliver essential services, including power, mobility, and connection \cite{b2}. 

\begingroup
\newlength{\xfigwd}
\setlength{\xfigwd}{\textwidth}
\begin{figure*}[ht!]
\centering
\includegraphics[width=7in]{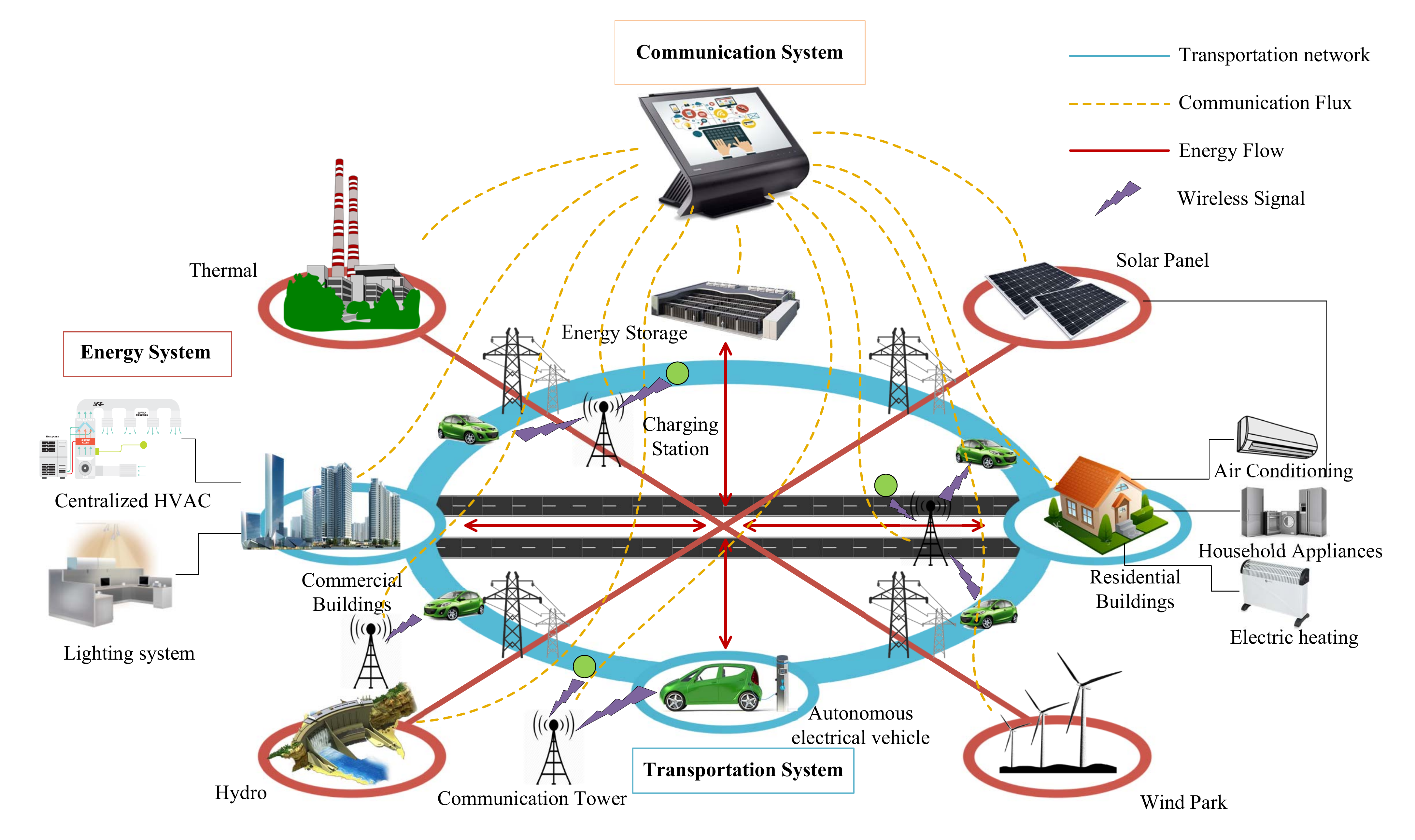}
\caption{A vision of different infrastructure systems in future connected communities.}
\label{fig:fig1}
\end{figure*}
\endgroup

Figure \ref{fig:fig1} shows our vision of future smart and connected communities.
These connected communities will include three key infrastructure systems: energy, transportation, and communication infrastructure.
The energy system includes different heterogeneous components such as increasing renewable energy resources, buildings as virtual batteries as well as massive adoption of electric vehicles (EVs).
The transportation system is also expected to undergo an unprecedented evolution due to the shift toward full electrification and autonomous vehicles \cite{b3,b4,b5}.
The communication system will become an indispensable enabler to support the aforementioned infrastructure systems, link various system components, and coordinate the operational sequences \cite{b6,bR14}. 

Infrastructure interdependency involves a bidirectional relationship between systems in which the state of each system is dependent on and intertwined with the other. 
For a simple example, communication networks need power from the electric grid to function, while the electric grid needs the communication networks to dispatch generation facilities according to the demands.
With these codependent relationships, disturbances and capacity stresses on one system can affect the other, potentially creating a cascading effect that compromises the systems' operations. 

Today, these interdependencies are most often felt in critical infrastructure systems (CIS) -- which include energy, transportation, and communication systems, among many others.
It is well proven that CIS are indeed highly intertwined and codependent \cite{bR1-7,bR2-8,bR3-9}.
In the future, increased electrification and connectivity will further complicate these interconnections, as demonstrated in Figure {\ref{fig:fig1}}. 
To date, interdepenedency modeling primarily focuses on averting potentially catastrophic failures and minimizing the risks of future failures. 
This is justly so; while these interdependencies today often go unnoticed in their typical operation, they effectively aggravate and compound system failures when subjected to various stresses (such as cyber attacks, congestion, and natural disasters). 
In the near future, it is likely that infrastructure vulnerabilities will become increasing present in day-to-day operations due to their deeply crosslinked nature. 

\begingroup
\setlength{\xfigwd}{\textwidth}
\begin{figure*}[ht]
\centering
\includegraphics[width=6in]{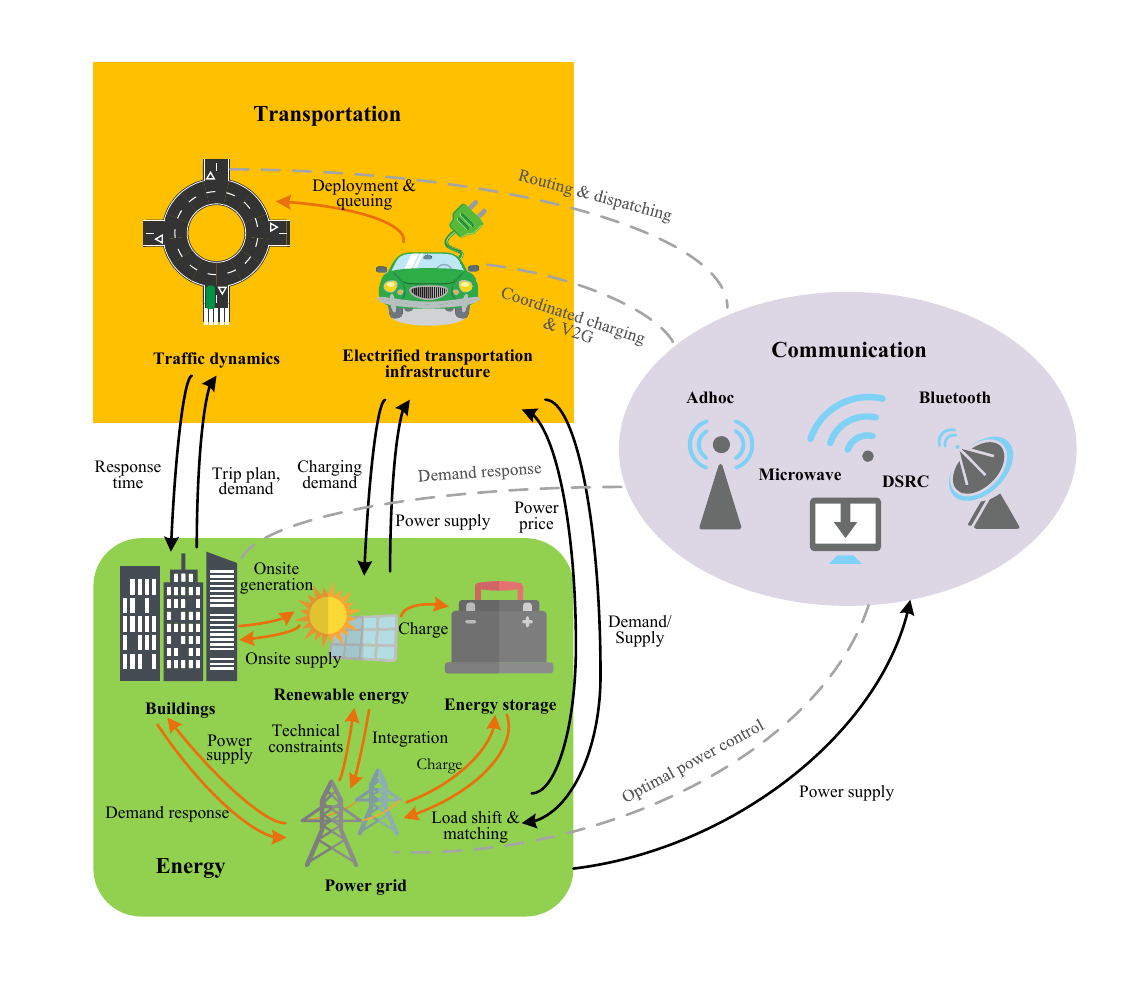}
\caption{A schematic of the interdependent infrastructure systems in connected communities.}
\label{fig:fig2}
\end{figure*}
\endgroup

Due to the inherent complexities in future SCC infrastructure, we reorganize the sophisticated and elusive relationships in Figure {\ref{fig:fig1}} into Figure {\ref{fig:fig2}}, which illustrates the interaction among the different systems considered.
The interdependencies exist both within and between systems and present themselves on several levels, which are commonly classified under four types: (1) physical, meaning the state of one system depends on the material output(s) of the other, and vice versa; (2) cyber, meaning there is information exchange between the systems; (3) geographic, meaning the infrastructure components are in close spatial proximity with each other; and (4) logical, meaning there is a different mechanism (e.g. policy, legal, or regulatory regime) that logically ties the infrastructure systems.
A physical interdependency example is in the electricity exchange between transportation and energy systems, where EVs take electricity from the grid while charging, but they are can provide ancillary services with their large storage capacities and quick discharging abilities \cite{bR4-10}.
Further, a logical interdepenency arises when the utility company increases the electricity price due to rising charging demand, which in turn affects the drivers' charging patterns and the traffic condition \cite{bR5-11}.
On the cyber level, the quality of the supporting communication services affects the transportation and energy systems (shown as the dashed lines). 
In the transportation system, communication services enable vehicle routing, dispatching, coordinated charging, and the implementation of vehicle to grid interaction. 
These interdependencies indicate a need to consider multiple infrastructure systems when operating future SCCs.
To this end, modeling and simulation is an effective method where these complex, multilevel interactions among several infrastructure systems can be studied \cite{bR6-12}.

\begin{table*} [hb]
\caption{State-of-the-art of integrated models for energy, transportation, and communication infrastructure.}
\label{table1}
\makebox[\linewidth]{
\begin{tabular} {|c|c|c|c|c|}
\hline
\multicolumn{5}{|c|}{\cellcolor{gray!25}\textbf{Energy \& Transportation}} \\
\hline
\multirow{2}{*}{\textbf{Model name}} &
\multicolumn{2}{c|}{\textbf{Simulation tools}} & 
\multirow{2}{*}{\textbf{Use cases}} &
\multirow{2}{*}{\textbf{Open Source}} \\ 
\cline{2-3} & \textbf{Energy} & \textbf{Transportation} & \multicolumn{1}{c|}{} & \multicolumn{1}{c|}{}\\ 
\hline
NA \cite{bR13} & \multicolumn{2}{c|}{In-house codes} & Power flow and travel cost optimization &Unknown \\
\hline
NA \cite{b23} & \multicolumn{2}{c|}{In-house codes} & Transportation-energy nexus assessment &Unknown \\
\hline
NA \cite{b24} & OpenDSS & DYNASMART & Impact of EVs at the distribution feeder level & Unknown\\
\hline
NA \cite{b37} & MATPower & \makecell{Clean Mobility\\ Simulator} & \makecell{Power system safety; transportation \\ performance assessment} & Full\\
\hline
NA \cite{b38} & DIgSILENT PowerFactory & MATSim, EVSim & Charging strategy assessment; V2G & Partial\\
\hline
NA \cite{b39} & EMTP-rv & SUMO & \makecell{Charging strategy assessment; Complex \\ network planning} & Partial\\
\hline
\multicolumn{5}{|c|}{\cellcolor{gray!25}\textbf{Energy \& Communication}} \\
\hline
\multirow{2}{*}{\textbf{Model name}} &
\multicolumn{2}{c|}{\textbf{Simulation tools}} & 
\multirow{2}{*}{\textbf{Use cases}} &
\multirow{2}{*}{\textbf{Open Source}} \\ 
\cline{2-3} & \textbf{Energy} & \textbf{Communication} & \multicolumn{1}{c|}{} & \multicolumn{1}{c|}{}\\ 
\hline
EPOCHS \cite{b25} & PSCAD, PSLF & ns-2 & Multi-agent protection and control systems & Partial\\
\hline
INSPIRE \cite{b26} & PowerFactory & OPNET & \makecell{Wide-area monitoring, protection, and \\ control (WAMPAC) in high voltage grid} & Partial\\
\hline
VirGIL \cite{b27} & \makecell{Energyplus, Dymola, \\ DigSILENT PowerFactory} & OMNeT++ &Demand response; Vol/var control & Partial\\
\hline
Gridspice \cite{b40} & GridLab-D, MATPOWER & Simplified &\makecell{Transimission and distribution system;  \\ Demand side management/Demand response} & Partial\\
\hline
NA \cite{b41} & PSCAD/EMTDC & OPNET & WAMPAC in high voltage grid & Partial\\
\hline
NA \cite{b42}& Simulink, JADE & OMNeT++ & WAMPAC in middle voltage grid & Partial\\
\hline
\multicolumn{5}{|c|}{\cellcolor{gray!25}\textbf{Transportation \& Communication} }\\
\hline
\multirow{2}{*}{\textbf{Model name}} &
\multicolumn{2}{c|}{\textbf{Simulation tools}} & 
\multirow{2}{*}{\textbf{Use cases}} &
\multirow{2}{*}{\textbf{Open Source}} \\ 
\cline{2-3} & \textbf{Transportation} & \textbf{Communication} & \multicolumn{1}{c|}{} & \multicolumn{1}{c|}{}\\ 
\hline
Veins \cite{b28}& SUMO & OMNeT++ &\makecell{Inter-vehicular communications\\ performance evaluation} & Full\\
\hline
VNS \cite{b29} & DIVERT 2.0 & ns-3 &\makecell{Inter-vehicular communications\\ performance evaluation} & Full\\
\hline
HINT \cite{b30} & SUMO & ns-3 &Traffic management scenario analysis & Full\\
\hline
VSimRTI \cite{b31} & SUMO/VISSIM & OMNeT++, ns-3 & Dynamic traffic flow simulation & Partial\\
\hline
iTETRIS \cite{b43} & SUMO & OMNeT++ & Largescale vehicular network simulation & Full\\
\hline
NA \cite{bR10, bR11}& \multicolumn{2}{c|}{In-house codes} & Vehicle routing algorithm analysis & Unknown \\
\hline
\multicolumn{5}{p{350pt}}{
\break
\textbf{NA}: Not applicable; \textbf{Partial}: Partial simulators are open source; \textbf{Full}: All the simulators are open source; 
\textbf{Unknown}: The open source status is unknown. }\\
\end{tabular}
}
\end{table*}

\subsection{Prior Works}
One cost-effective way to evaluate the integrated infrastructure systems in a connected community is through computer-aided modeling.
The existing community modeling approaches can be grouped into two types: discipline-specific modeling and integrated modeling.
The discipline-specific modeling of community infrastructures has been well-developed over the past decades and numerically dedicated to particular purposes of urban building energy modeling \cite{b8,b9,b10,b11}, urban mobility modeling \cite{b12,b13,b14,b15,b16}, and communication network simulation \cite{b17,b18,bR12}.
Urban building energy models are targeted at predicting the energy use in a community, which is a relatively nascent field \cite{b8}. 
For the urban mobility simulation, Multi-Agent Transport Simulation (MATSim), Simulation of Urban Mobility (SUMO), and Transportation Analysis and Simulation System are the three mainstream, open source simulators to conduct transportation analysis and dynamic traffic assignment \cite{b19}.
In the context of communication network modeling in the interconnected infrastructure systems, ns2/ns3 \cite{b17} and OMNeT++ \cite{b18} are widely used due to their open source nature.
Although the discipline-specific modeling can provide accurate and efficient results for the phenomenon of interest, its functionality deteriorates when it comes to the interdependent infrastructure systems due to significant simplifications \cite{b20}. 

In contrast, integrated modeling approaches are more suited for the interdependent infrastructure systems. 
Several modeling techniques have been adopted, including agent-based modeling, input-output models, network based approaches, and system dynamics.
Some recent works \cite{b21,b22} considered mathematical modeling of interdependent infrastructure; however, these simplified numerical simulations have limited capabilities in modeling real-world SCCs.
To achieve a higher fidelity, specialized software packages that use validated model libraries and tailor-made solvers are coupled by a certain orchestration mechanism.
Table \ref{table1} shows a non-exhaustive list of the references that model energy, transportation, and communication systems and their interdependencies.
Details in terms of the platforms, simulators, use cases, and open source statuses are also included.

For SCC applications, most of the related works have focused on the integration of two systems and utilized proprietary integrated modeling tools.
Coupling the energy and transportation systems, Farid \cite{b23} proposed a modeling framework using graph theory and petri-net models to assess the effects of transportation-electrification.
Su et al. \cite{b24} coupled the OpenDSS and DYNASMART to evaluate the potential impacts of electric vehicle charging at the feeder level.
However, these studies do not simulate the detailed communication processes and control management, in which the energy, transportation, and communication systems are coupled.
For the energy and communication systems, profound efforts have been made to couple continuous power system simulators with discrete communication network simulators.
The typical simulation platforms include the electric power and communication synchronizing simulator (EPOCHS) \cite{b25} and the integrated co-simulation of power and information and communications technology (ICT) systems for real-time evaluation (INSPIRE) \cite{b26}.
A recent coupled platform built by Chatzivasileiadis et al. \cite{b27} implemented the Ptolemy II framework to integrate the building models, power system software, and the communication network software. 
The platform can be used to investigate demand-response strategies and voltage control strategies. 
For the transportation and communication systems, a traffic simulator is used to understand traffic patterns and to reduce real-life traffic congestion. 
Conversely, network simulators concentrate on network states and events such as routers, mobile nodes, and transmission rates. 
The combined modeling of transportation and communication systems can be utilized in the analysis of inter-vehicular communication performance \cite{b28,b29}, traffic management scenario \cite{b30, bR13}, and dynamic traffic flow \cite{b31}.
\newpage
While most existing research focuses on integrating two systems, Bedogni et al. \cite{b32} integrated three systems (a mobility simulator, a dynamic power distribution network simulator, and a UMTS communication network simulator). 
However, the scalability of their coupled model is limited, and their proposed model is dedicated to the power control of EV charging stations. 
The rigid coupling approach adopted in their study blocks the model itself from adapting other scenarios that are common in the context of SCCs infrastructure systems.

Beyond SCC applications, the state-of-the-art approaches in CIS modeling are well summarized in the existing literature \cite{bR3-9,bR7-34}.
Many successful projects have modeled the interdependencies among CIS.
For example, several groups \cite{bR4,bR5,bR6,bR7} proposed different cascading failure models for interdependent power and communication networks. 
Similarly, Heracleous et al. \cite{bR8} developed hybrid automata models for the interdependent power, water, and communication infrastructures.
While related, interdependency modeling for CIS may not be extensible to study SCCs during typical operation.
Interdependent CIS models are often built to understand how these complex, interdependent systems will respond to disruptions and changing conditions.
As such, their simulation goal is not to produce an ``exact'' outcome, but to illustrate possible outcomes to inform urban planning and design.
By contrast, the underlying dynamics and model fidelity are crucial in the SCC modeling case, where the intention is to study use cases for typical operation.

The above literature review demonstrates that discipline-specific modeling for SCCs usually specializes in an isolated system and simplifies or even neglects the interdependencies among other systems. 
Meanwhile, integrated modeling does consider the interdependencies; however, the existing models are often limited by their proprietary nature and two-system focus.
Furthermore, interdependency modeling in CIS applications is well developed but may not be fully extendable to studying SCC operational cases.
Therefore, it is important to have an open source modeling framework that can cover different operational scenarios of the entire infrastructure. 
This is particularity important because the reactions to events in a single system may spread across multiple systems due to the interdependencies present \cite{b33}.

\subsection{Contributions}
In response to the aforementioned limitations of current modeling of infrastructure systems, the main contribution of this paper is to develop an open source, flexible, and extensible modeling framework for studying interdependent energy, transportation, and communication infrastructures during typical SCC operation. 
The open source model can be found at https://www.colorado.edu/lab/sbs/scc-library.
We propose a novel multi-level, multi-layer, multi-agent approach to enable flexible modeling of the three interconnected systems of energy, transportation, and communication. 
The proposed framework is then implemented on Modelica-based modeling platforms \cite{b34,b35,b36}, which provides important features like objected-oriented, acausal modeling and equation-based schemes. 
In addition, we develop various component and system-level models for energy, transportation and communication systems. 
The objective of these models is to demonstrate the application of the proposed framework; therefore, the interdependency and system models are selected to adequately test the framework while minimizing complexities at the current stage.
To the best of our knowledge, this is one of the first attempts to couple all the three energy, transportation, and communication systems in a flexible and scalable way to evaluate and quantify the underlying interdependencies of their infrastructures for SCCs. 

The rest of the paper is organized as follows. 
Section~\ref{section:3M} presents the proposed smart community modeling framework based on the multi-level, multi-layer, multi-agent (3M) approach for coupling multiple infrastructure systems.
Section~\ref{section:ModDes} introduces the Modelica implementation of the proposed modeling framework as well as various component and system models in Modelica. 
Section~\ref{section:CasStu} presents three case studies to demonstrate the application scenarios of the modeling framework and the measurable interrelations among the three systems.
Finally, concluding remarks are presented in Section~\ref{section:Con}.

\section{Multi-layer, Multi-block, Multi-agent Approach} \label{section:3M}
This study proposes a 3M approach for the flexible and extensible modeling of the coupled systems in a community level. 
This generalized modeling approach can be adapted to different scenarios and use cases in the design and operation of SCCs. 
The modular nature of the 3M approach allows various system combinations to be studied (two-system models, three-system model, and so forth).
Furthermore, the generalized, adaptable framework can readily accept a variety of other component models along with their dependency interconnections to evaluate many dynamic systems.
In this section, we use the three-coupled system to illustrate the modeling approach.

Figure \ref{fig:fig3} illustrates the principle of the proposed 3M approach. 
The multi-layer represents a hierarchical structure, which consists of a community layer, a block layer, and a system agent layer from the top level to the bottom. 
At the community layer, the entire community will be divided into several functional blocks, such as residential blocks, commercial blocks, industrial blocks, or mixed-functional blocks. 
It could also be divided according to domain-specific entities, such as distinct power distribution networks. 
The block layer has three system agents for the energy, transportation, and communication systems. Different infrastructure systems should be mapped accordingly in one block. 
Within each system agent (system agent level), there will be individual subagents for infrastructure nodes. 
For instance, an energy system agent could have subagents for renewable generation, batteries, and utility customers. 
Likewise, the transportation system agent could have subagents for connected roads, charging stations, and vehicles. 
The communication system agent could consist of different communication devices (control centers, backbone routers) according to the different types of communication networks (e.g. cellular/LTE, WiFi, or PLC). 

\begingroup
\setlength{\xfigwd}{\textwidth}
\begin{figure*}[ht]
\centering
\includegraphics[width=6.8in]{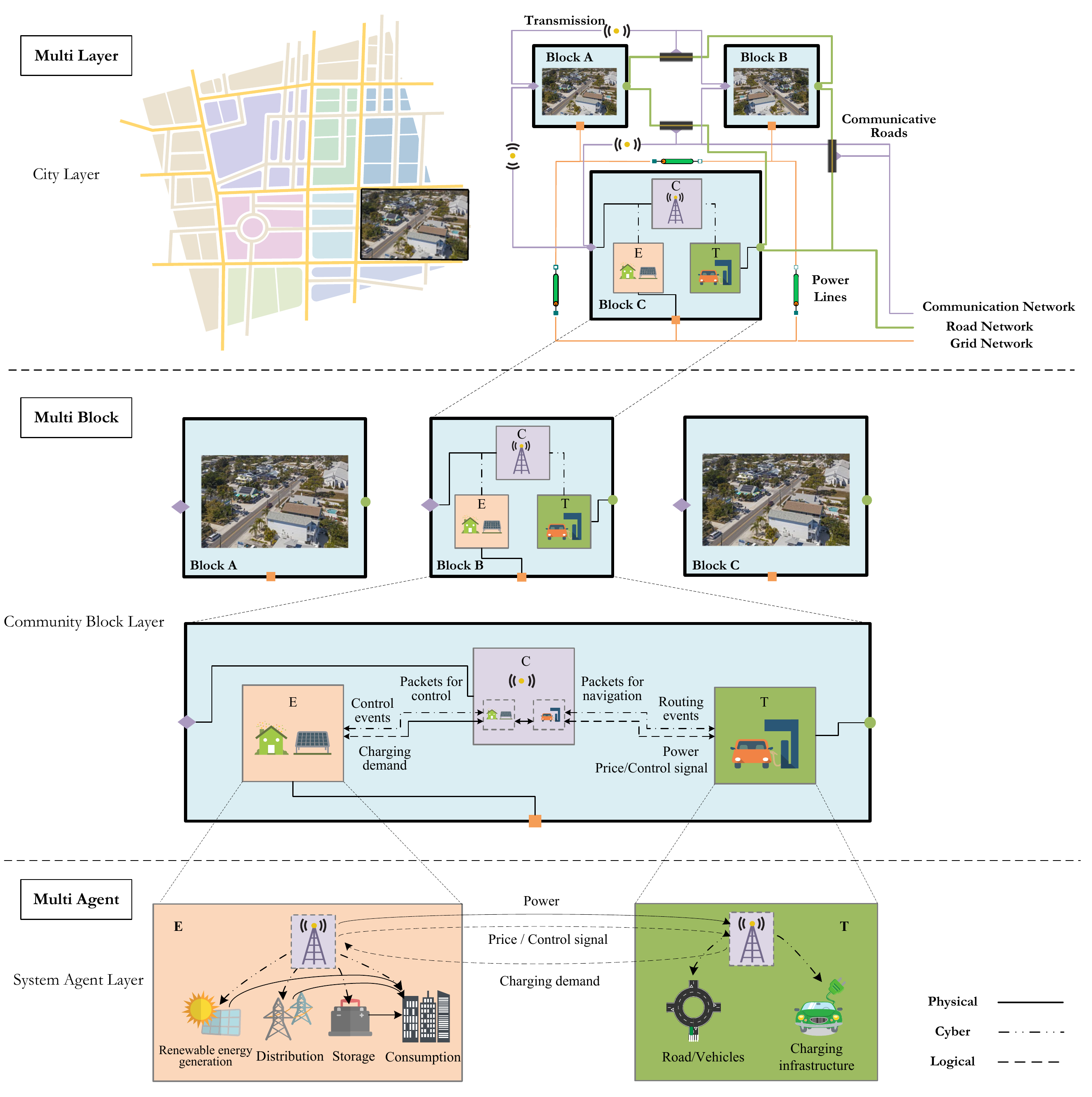}
\caption{A schematic of the 3M approach for modeling coupled systems.}
\label{fig:fig3}
\end{figure*}
\endgroup

We next illustrate the interdependencies and data exchange between these agents and blocks in a bottom-up fashion. 
At the system agent layer, multiple agents could exchange data through certain interfaces.
As shown in Figure {\ref{fig:fig3}}, the interdependencies can be physical, cyber, or logical, and both internal and external dependencies are present.
In the energy domain, the different components will communicate with each other to achieve stable and economic operations of the power grid. 
For example, in a demand response application, the electricity demand of each customer can be properly rescheduled via the communication network to reduce the peak demand between the power grid and its customers. 
While in the transportation domain, different components will also interact with each other to achieve a good traffic condition. 
For example, the queuing and power availability information in the charging stations and the routing information could be broadcast to the vehicles for optimal routing and charging arrangements. 
The aforementioned connections are the inner data exchanges in each domain. 
To ensure a collective optimal operation of the energy and transportation domains, the agents from the energy system will calculate the electricity price under different load profiles and trigger the necessary power controls of the charging stations. 
In response to this, the transportation agents will calculate the traffic condition, decide the routing and charging behavior of electric vehicles, and determine the charging load of each charging station node. 
In this system agent layer, we could select the needed component model according to the use cases. 

At the block layer, we encapsulate all the component models from the system agent layer and merge the different communication centers into one communication system agent, which would calculate the communication latency among the different components. 
The generalized data exchange between different system agents are propagated from the data exchange in the system agent layer. 
The energy system agent will transfer power from the grid to charge the EVs (physical) and send the packets for the power system control through the communication system agent (cyber). 
The communication simulators within the communication system agent will then return the control events (cyber) to the energy system agent. 
Likewise, the transportation system agent will send the packets for navigation purpose to the communication system agent (cyber), and it will receive the routing events from the communication system agent (cyber). 
Lastly, the energy system and transportation system agents will exchange the control/price signal and the charging demand via the communication agent (logical). 
Note that the aforementioned generalized dependencies could be specialized by different use cases, which should be aligned with the system agent layer.

In the community layer, the different blocks are linked with each other through the corresponding ports.
The power port links the energy system agents in different blocks to a power grid network, and the traffic port links the transportation system agents in different blocks to a road network. 
The communication system agents will communicate in a larger network, and the information will be collected and scattered in a large scale communication center or in distributed hubs. 
In future SCCs, the roads and the power lines could also be a communicative component in this layer. 

The proposed 3M approach allows the flexible and extensible modeling of multiple infrastructures. 
The modeler could save efforts by using this standardized and generalized approach since the interfaces and the data exchange ports from the upper layer are fixed. 
As such, only the component models in the lower layer require designing and refinement. 
The proposed modeling framework facilitates the model reuse and can be used to investigate different use cases in the design and operation of SCCs. 
In the next section, we will introduce our efforts to designing and modeling preliminary multi-infrastructure systems in order to test and validate the 3M approach.

\section{Model Description and Implementation} \label{section:ModDes}
\label{sec:guidelines}
Using the proposed 3M approach, we implement the coupled models in Modelica, an object-oriented modeling language.
Section \ref{section:SAM} illustrates the implementation of system agent models of the energy, transportation, and communication systems. 
Sections \ref{section:BLM} and \ref{section:CLM} discuss the implementation of block layer and community layer models, respectively.

\newpage
\subsection{System Agent Model} \label{section:SAM}
\subsubsection{Energy System} \label{energysystem}
The energy system is composed of renewable generation equipment (e.g. photovoltaic (PV) panels and wind turbines), energy storage devices, power draw from the grid, and distribution systems. 
The feeders of the distribution system connect various types of loads, such as the load from buildings, EVs, communication towers, and general loads in the power grid. 
Figure \ref{fig:fig4} shows the Modelica implementation of the energy system. The following subsections describe the detailed description and implementation of different energy component models.

\begingroup
\setlength{\xfigwd}{\textwidth}
\begin{figure}[htbp]
\centering
\includegraphics[width=3.5in]{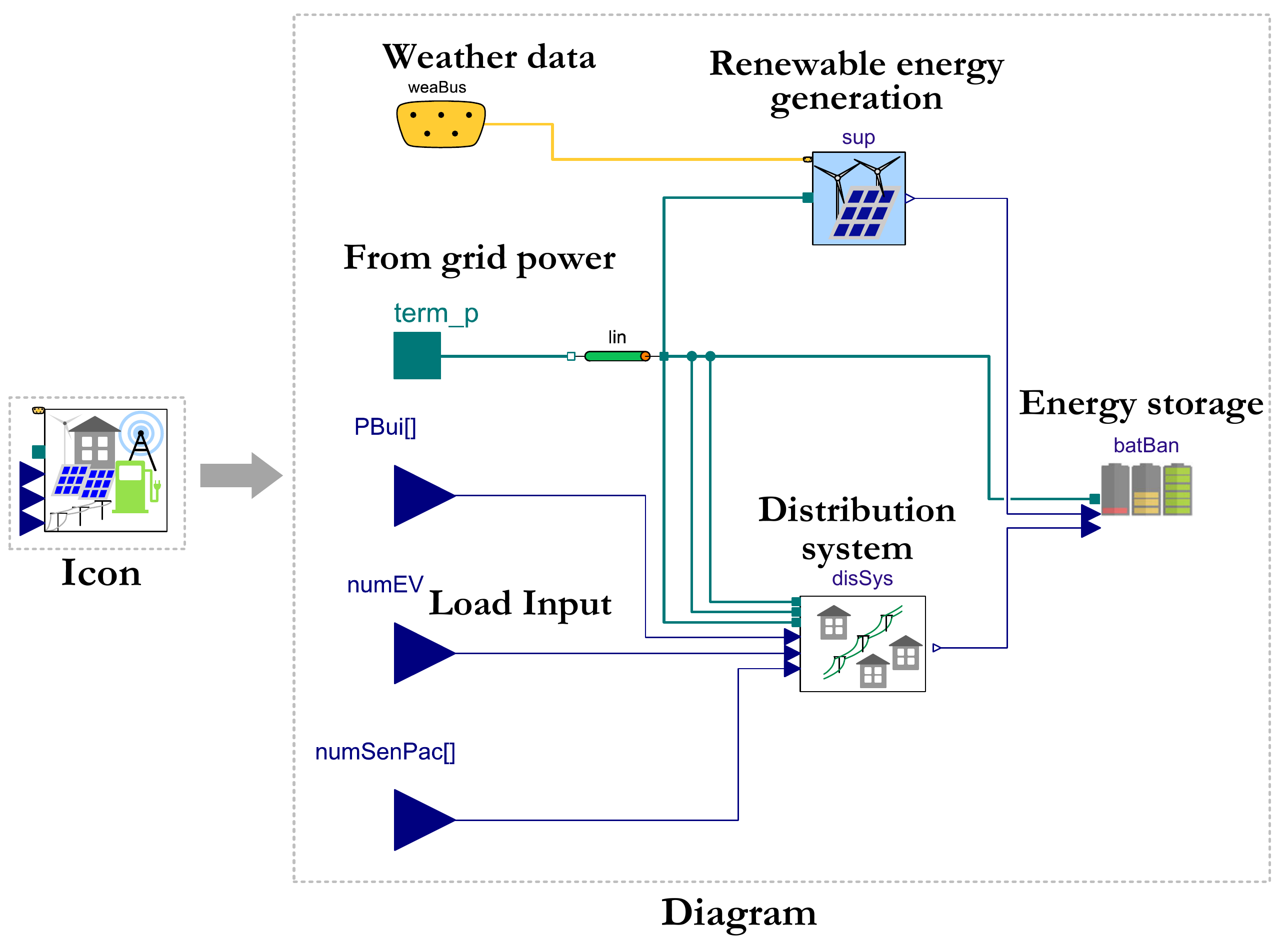}
\caption{Implementation of the energy system models in Modelica.}
\label{fig:fig4}
\end{figure}
\endgroup

\paragraph{Renewable Energy Generation}
{In this paper, we assume renewable energy generation consists of energy supplied from  PV and wind turbines. 
The PV system is modeled with the $PVSimpleOriented$ model in the Modelica Buildings Library. 
The total aggregated electrical power $P_\textrm{PV}$ generated by the PV systems is represented by:
\begin{equation}P_\textrm{PV}= \sum_{k=1}^{N_\textrm{PV}}{(A_{k} \cdot f_{\textrm{act,}k} \cdot \eta_{k} \cdot G_{k} \cdot \eta_{\textrm{DCAC,}k})}
\label{eq:eq1}\end{equation}
where $N_\textrm{PV}$ is the total number of PV arrays, $A$ is the area of each PV array, $f_{\textrm{act}}$ is the fraction of the aperture area, $\eta$ is the PV efficiency, $G$ is the total solar irradiation, and $\eta_{\textrm{DCAC}}$ is the efficiency of the conversion between direct current (DC) and alternating current (AC). 
In this model, $G$ is the sum of direct irradiation $G_{\textrm{Dir}}$ and diffuse irradiation $G_{\textrm{Dif}}$:
\begin{equation}G=G_{\textrm{Dir}}+G_{\textrm{Dif}}\label{eq:eq2}\end{equation}
\begin{equation}G_{\textrm{Dir}}=\max(0, \cos (\theta) \cdot H_{\textrm{DirNor}})\label{eq:eq3}\end{equation}
\begin{equation}G_{\textrm{Dif}}=G_{\textrm{SkyDif}}+G_{\textrm{GroDif}}\label{eq:eq4}\end{equation}
where $\theta$ is the solar incidence angle on the surface, and $H_{\textrm{DirNor}}$ is the direct normal radiation. 
$G_{\textrm{SkyDif}}$ and $G_{\textrm{GroDif}}$ are the hemispherical diffuse solar irradiation on a tilted surface from the sky and the ground, respectively.

The $WindTurbine$ model in the Buildings library is adopted for wind turbines. 
For a single turbine, the model computes generated power $P_t$ as a function of the wind speed. 
The aggregated active electrical power generated by all the wind turbines $P_\textrm{win}$ is calculated as:
\begin{equation}P_\textrm{win}=\sum_{k=1}^{N_\textrm{win}}{(P_{t,k} \cdot scale_{k} \cdot \eta_{\textrm{DCAC\textunderscore}win,k})}
\label{eq:eq5}\end{equation}
where $scale$ is used to scale the wind power generation based on the $P_t$; $N_\textrm{win}$ is the total number of wind turbines; and  $\eta_{\textrm{DCAC\textunderscore win}}$ is the rate of conversion from DC to AC.
}

\paragraph{Energy Storage}
{A new energy storage model is built based on the \textit{Electrical.AC.OnePhase.Sources.Battery} model in the Buildings library, since the base model does not enforce that the state of charge is between zero and one.
The new model provides a control sequence such that only a reasonable amount of power is exchanged. 
The new model and corresponding control sequence are shown in Figure \ref{fig:fig5}. 
The control logic here is that the power generated by the renewable system covers the demand load in the first place. 
When there is excessive power, the battery starts to charge. When the power demand exceeds a certain threshold, the battery starts to discharge. 
In other cases, the battery remains stand-by.}

\begingroup
\setlength{\xfigwd}{\textwidth}
\begin{figure}[h]
\centering
\includegraphics[width=3.5in]{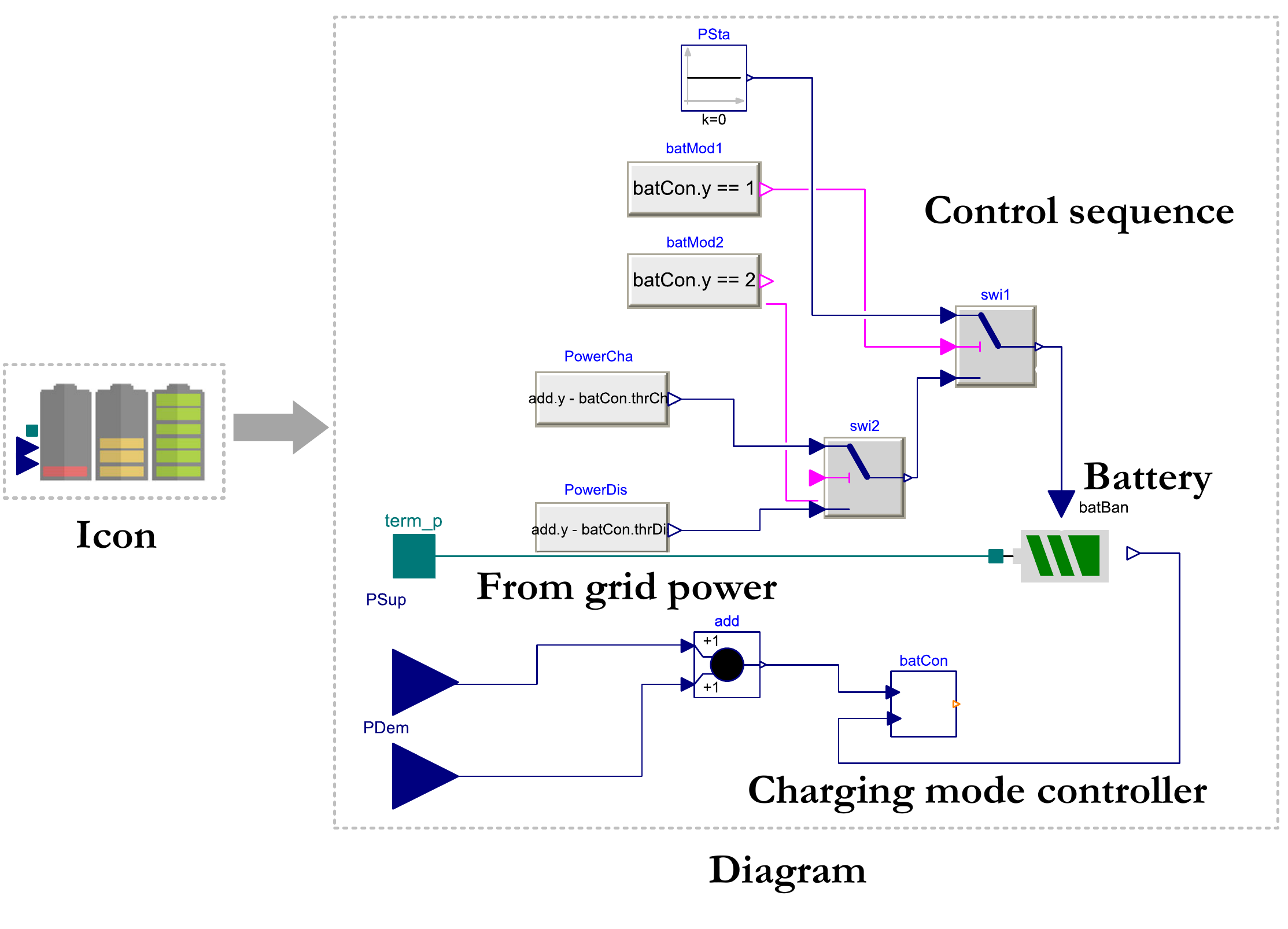}
\caption{Diagram of Modelica models for energy storage and its control logic.}
\label{fig:fig5}
\end{figure}
\endgroup

\paragraph{Power Distribution Network}
{Different configurations of power distribution networks are built in the energy system agent model.
These networks can be used to investigate the performance of power distribution systems. 
Figure \ref{fig:fig6} shows the IEEE 16-node test feeder configuration \cite{b44} and its implementation in Modelica.

\begingroup
\setlength{\xfigwd}{\textwidth}
\begin{figure}[ht]
\centering
\includegraphics[width=3.5in]{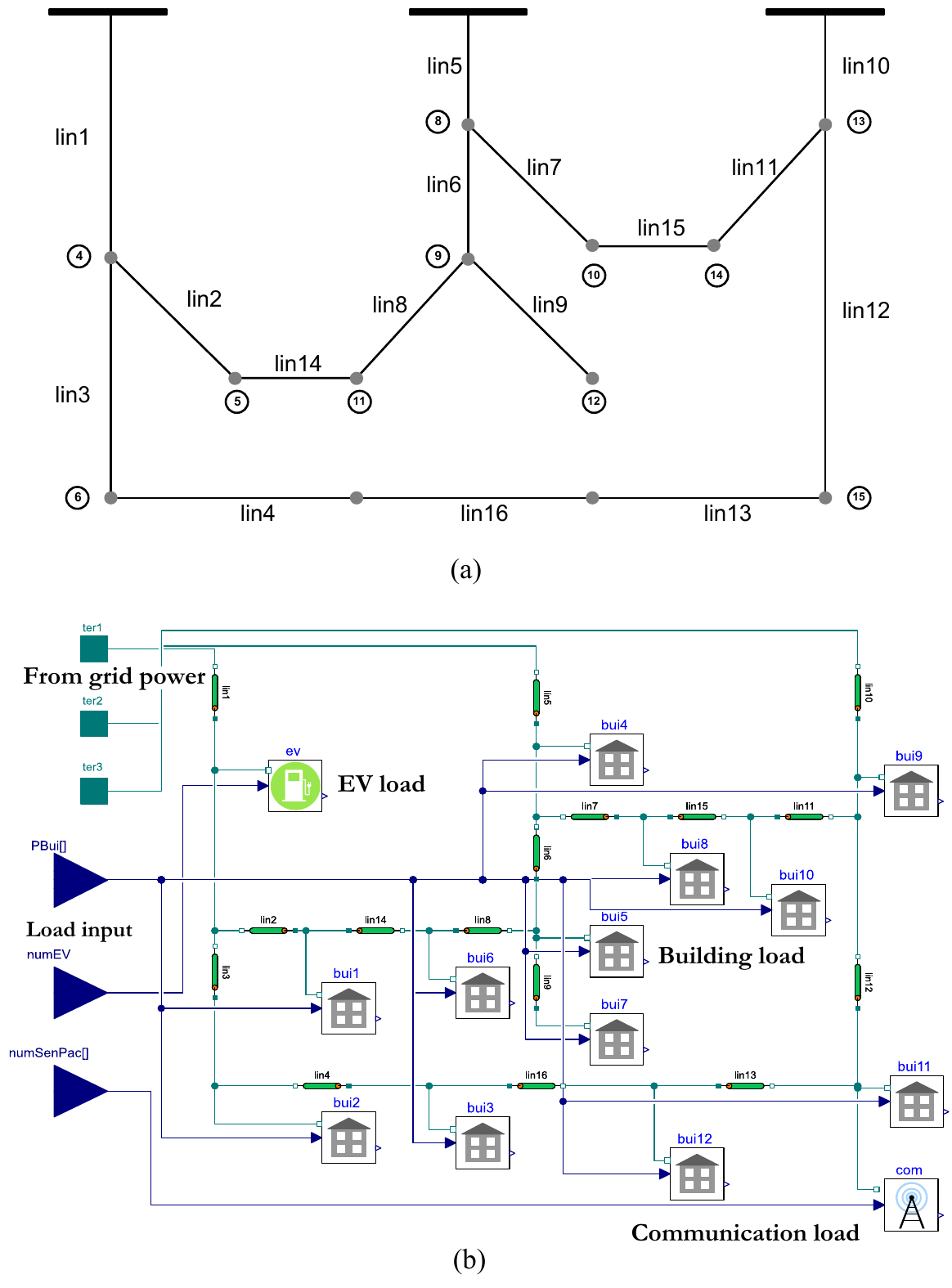}
\caption{The IEEE 16-node test feeder configuration: (a) System schematic and (b) Modelica Implementation.}
\label{fig:fig6}
\end{figure}
\endgroup

Different types of loads (buildings, EVs, general, and communication components) are connected to the feeders. 
The loads from buildings can be calculated using grey-box models such as RC models \cite{b45} or imported by data-driven models \cite{b46}. 
The aggregated EV charging power $P_\textrm{EV}$ is calculated as follows:
\begin{equation}P_\textrm{EV}=\sum_{k=1}^{N_\textrm{char}}  P_{\textrm{char,}k}
\label{eq:eqr8}\end{equation}
where $N_\textrm{char}$ is the total number of the charging EVs in equation (\ref{eq:eq10}), and $P_{\textrm{char},k}$ is the charging power of each EV. 
As a demonstration case, the current paper assumes $P_{\textrm{char}}$ is constant for all EVs \cite{b47}. 
The total load from all of the communication towers $P_\textrm{com}$ is given by \cite{b48}:
\begin{equation}P_{\textrm{com}}=\sum_{k=1}^{N_{\textrm{com}}}{(2Q_{c,k} \cdot E_{\textrm{elec,}k} + Q_{c,k} \cdot \epsilon_{\textrm{elec,}k} \cdot d^{\alpha}_{k})}
\label{eq:eq6}\end{equation}
where $N_\textrm{com}$ is the number of communication towers, $E_{\textrm{elec}}$ is the equipment power for sending and receiving packets, $Q_c$ is the packet throughput, $d$ is the distance of the transmission, and $\alpha$ and $\epsilon_{\textrm{elec}}$ are transmission coefficients. 
}

\paragraph{Power Draw from the Grid}
We use the \textit{Electrical.AC.OnePhase.Sources.Grid} model in the Buildings library to create the physics-based model for the utility supplied power \cite{b49}. 
The input for this model is a fixed voltage signal while the output is the power supplied by the utility to the power distribution system. 
The convention is that the power is positive if real power is consumed from the grid and negative if power flows back into the grid.

\subsubsection{Transportation System}\label{transport}
Traffic flow is intrinsically non-reproducible; it is therefore impossible to predict precise vehicle trajectories via models. 
However, it is recognized that the prediction of large-scale field quantities can be possible \cite{b50}. 
The traffic simulation models can be micro-, meso- and macroscopic with different granularities \cite{b51}. 
For example, macroscopic models describe the traffic as flows, velocities, and densities of vehicles, while microscopic models simulate individual vehicles down to basic physical and kinematic properties such as speed, locations, and fuel. 
In this research, we simulate transportation infrastructure from a macroscopic perspective. 
The road and charging station models are detailed in this section. 

\paragraph{Road Model}
The road model described here is able to model traffic outflow given the traffic inflow profile of a road, which utilizes the empirical flow-velocity correlation from literature \cite{b52,b53}, as shown in equation (\ref{eq:eq7}):
\begin{equation}
\begin{cases}
    U_{\textrm{ave}}&=\frac {\alpha_1 \cdot U_s} {1+{(\frac{V_{\textrm{ave}}}{C_t})}^{\beta}}\\
   \beta&=\alpha_2 + \alpha_3 \cdot {(\frac{V_{\textrm{ave}}}{C_t})}^3 
  \end{cases}\label{eq:eq7}\end{equation}
where $U_\textrm{ave}$ is the average road velocity; $V_{\textrm{ave}}$  is the average vehicle flow; $C_t$ is the road capacity; $\alpha_1$, $\alpha_2$, $\alpha_3$, and $\beta$ are regression parameters; and $U_s$ is the speed designed for the road. 
The traffic of the road can be measured by the traffic load {$\frac{V_{\textrm{ave}}}{C_t}$}.
Figure \ref{fig:fig7} demonstrates the relationship between $U_{\textrm{ave}}$ and $\frac{V_{\textrm{ave}}}{C_t}$ of different $U_s$ limits under a certain road type. 
We can then find that the velocity descends faster when traffic flow exceeds the road capacity $(\frac{V_{\textrm{ave}}}{C_t} > 1)$. 

\begingroup
\setlength{\xfigwd}{\textwidth}
\begin{figure}[ht]
\centering
\includegraphics[width=3.3in]{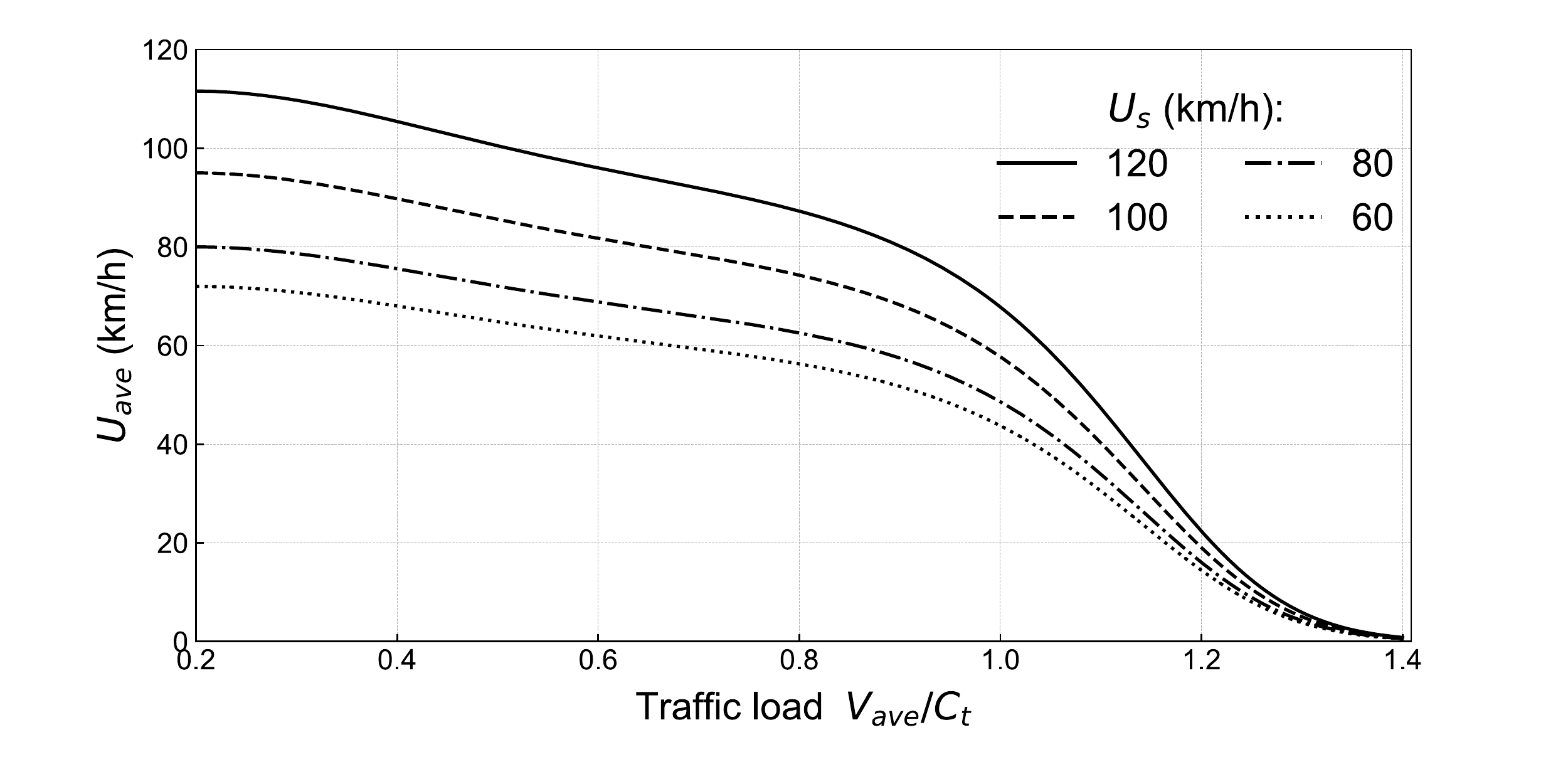}
\caption{Empirical model of traffic flow-velocity correlation for a certain road \cite{b52}.}
\label{fig:fig7}
\end{figure}
\endgroup

Equation (\ref{eq:eq8}) represents the relationship between the velocity, density, and traffic flow: 
\begin{equation}
    V_{\textrm{ave}}=\frac{U_{\textrm{ave}} \cdot \int (q_{\textrm{out}}-q_{\textrm{in}})\dd{t}}{L}
 \label{eq:eq8}\end{equation}
where $q_{\textrm{out}}$ and $q_{\textrm{in}}$ are respectively the traffic outflow and traffic inflow rates, and $L$ is the length of the road.

Equation (\ref{eq:eqr9}) represents the travel time on the road: 
\begin{equation}
    t_{\textrm{w/oCom}}=\frac{L}{U_{\textrm{ave}}}
 \label{eq:eqr9}\end{equation}
where $t_{\textrm{w/oCom}}$ is the travel time of the road without considering the communication system.
Combining equations (\ref{eq:eq7}-\ref{eq:eqr9}), the traffic condition on the road can be jointly obtained.
In our Modelica implementation, the road model consists of the $traTim$ model and the $varDel$ model (see Figure \ref{fig:fig8}). 
The pseudo-code of the $traTim$ model using the aforementioned principle is shown in Figure \ref{fig:fig9}. 
Based on the aforementioned principles, the $traTim$ model is built to calculate the travel time of the road section. 
The $varDel$ model is a delay block to mimic the travel delay time due to traffic conditions. 
Different road properties could be selected from the road type record datasets.
\begingroup
\setlength{\xfigwd}{\textwidth}
\begin{figure}[htbp]
\centering
\includegraphics[width=3.45in]{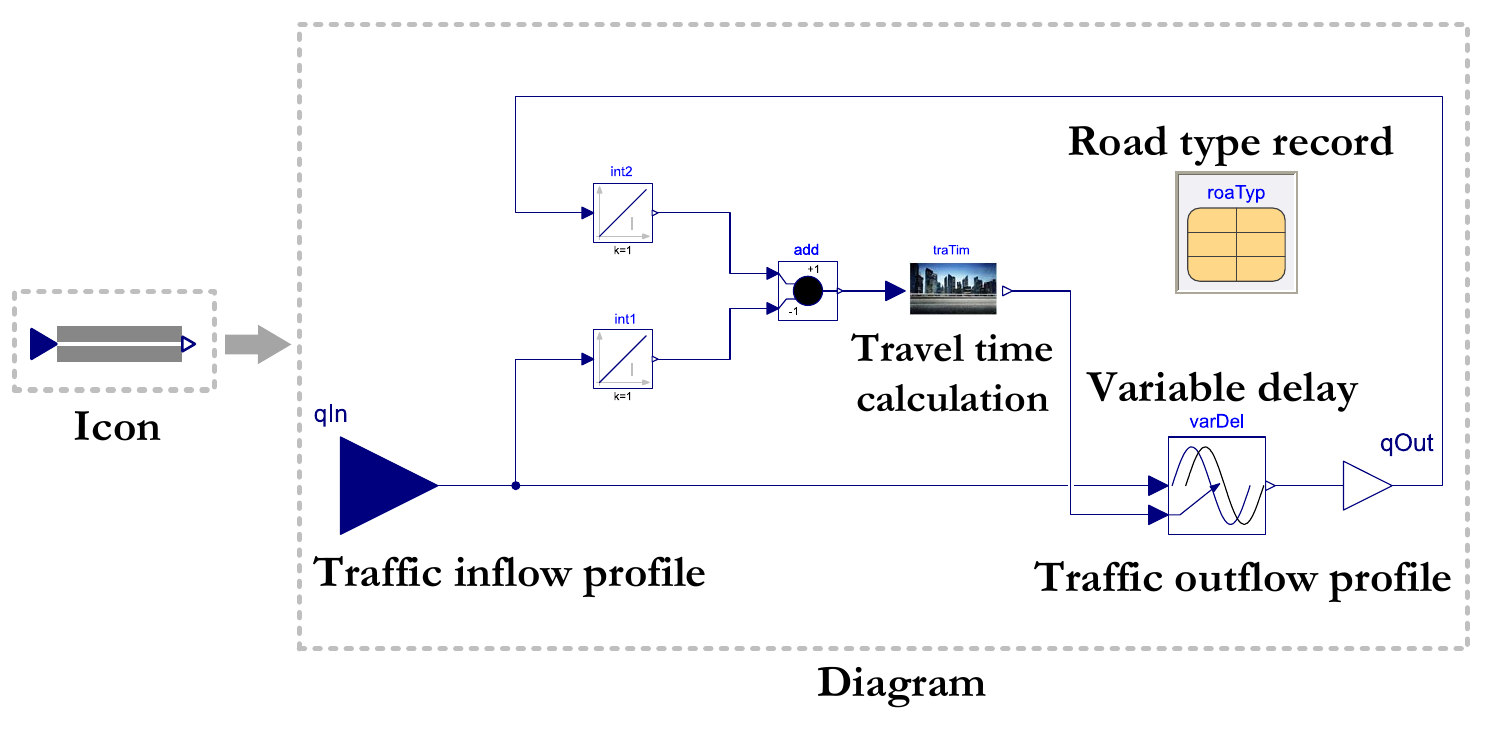}
\caption{Diagram of Modelica models for roads.}
\label{fig:fig8}
\end{figure}
\endgroup
\begingroup
\setlength{\xfigwd}{\textwidth}
\begin{figure}[htbp]
\centering
\includegraphics[width=3.4in]{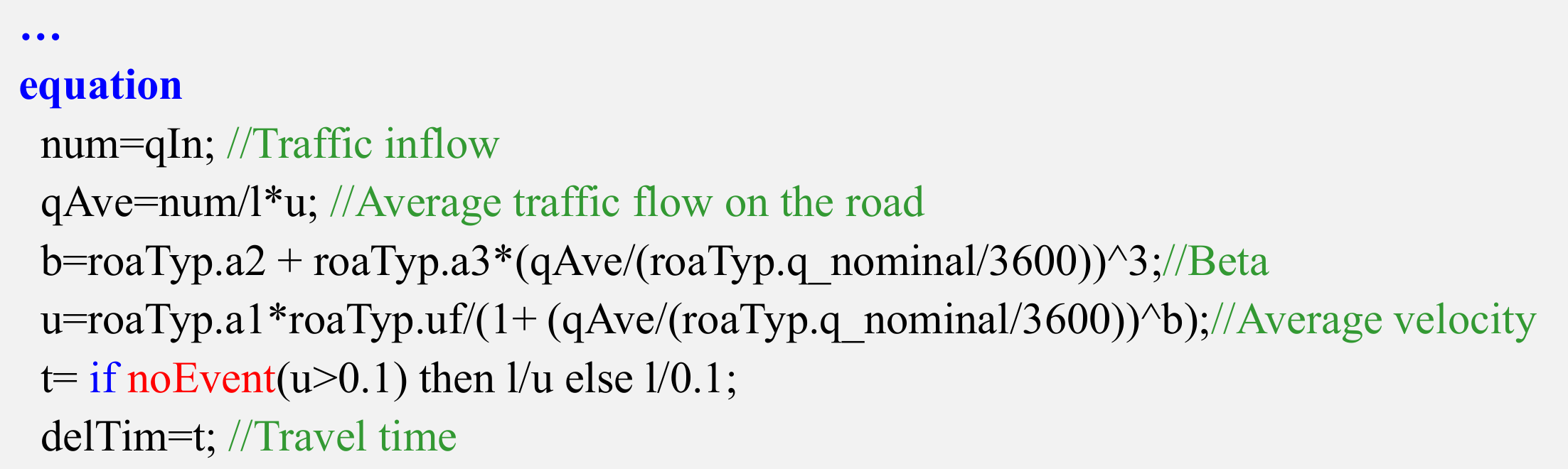}
\caption{Pseudo-code of $traTim$ model in Modelica.}
\label{fig:fig9}
\end{figure}
\endgroup

\paragraph {Charging EV Number Model}
The number of EVs in the community is calculated using a traffic flow balance:
\begin{equation}
   \dot N=\sum_{k=1}^{m_{\textrm{in}}} q_{\textrm{in},k}-\sum_{k=1}^{m_{\textrm{out}}} q_{\textrm{out},k}
 \label{eq:eq9}\end{equation}
where $\dot N$ represents the change of the number of EVs parked in the block; $m_{\textrm{in}}$ and $m_{\textrm{out}}$ denote the number of inlets and outlets of the station, respectively; and $q_{\textrm{in}}$ and $q_{\textrm{out}}$ represent the traffic inflow and outflow rates, respectively. 
While we recognize the number of vehicles is an integer in reality, we assume $N$ is continuous in this preliminary implementation.
The pseudo-code of the model using the aforementioned principle is shown in Figure \ref{fig:fig10}.
\begingroup
\setlength{\xfigwd}{\textwidth}
\begin{figure}[htbp]
\centering
\includegraphics[width=2.4in]{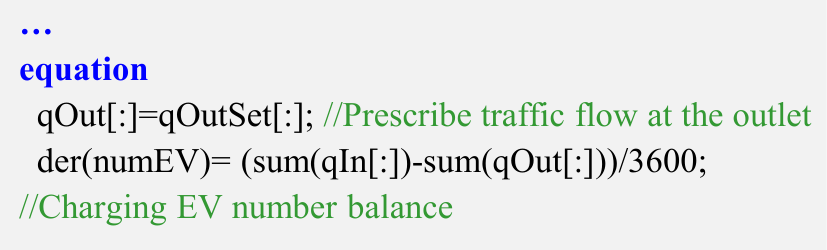}
\caption{Modelica pseudo-code for the number of charging EVs.}
\label{fig:fig10}
\end{figure}
\endgroup

There are many ways we can estimate the EV charging in the block level.
In this case study, we use a simplified probabilistic model.
For one single EV, its probability of charging at the block at a given hour is $p_i$, which can be obtained by survey data. 
We assume the number of the charging EVs at certain time represents a Poisson distribution. 
Therefore, we expect:
\begin{equation}
N_\textrm{char} = E(M) = p_i \cdot N
 \label{eq:eq10}\end{equation}
where $N_\textrm{char}$ is the total number of the charging vehicles, as discussed in equation (\ref{eq:eqr8}), and $N$ is number of parked vehicles in the block per equation (\ref{eq:eq9}).
For later applications, this simplified model can be replaced with more accurate and realistic models, such as Monte Carlo models \cite{b54}.

\subsubsection{Communication System} \label{communi}
In this preliminary case study, a simplified packet loss model is used to describe the transmission process in a wireless communication network \cite{b55}. 
The empirical relationship between the packet loss rate $\gamma$ and the normalized throughput $Q_c$ is:
\begin{equation}
\gamma=\kappa \cdot \sqrt{Q_c-C_c}
 \label{eq:eq11}\end{equation}
where $\kappa$ is a proportional coefficient and $C_c$ is the threshold of the transmission. 
We assume that $\gamma$ is directly proportional to $Q_c$ under certain bandwidth in the communication system. 
In this simplified model, we assume the transmission delay $Del$ equals 0 that there are negligible impacts from the re-transmission mechanism if a message is lost.

\subsubsection{Component Validation }
We validated the models using comparative testing and analytical verification. 
For example, comparative testing is used to validate the road model, which compares the Modelica simulation results with the data mentioned in literature \cite{b56}; this includes the flow rate data from the inlet to the outlet marked at 15 minute intervals for 6 hours. 
The detailed comparison can be seen in Figure \ref{fig:fig11}. 
From this, we can conclude that the model fits well with the literature data and could represent the traffic condition for the integrated model. 

\begingroup
\setlength{\xfigwd}{\textwidth}
\begin{figure}[h]
\centering
\includegraphics[width=3.3in]{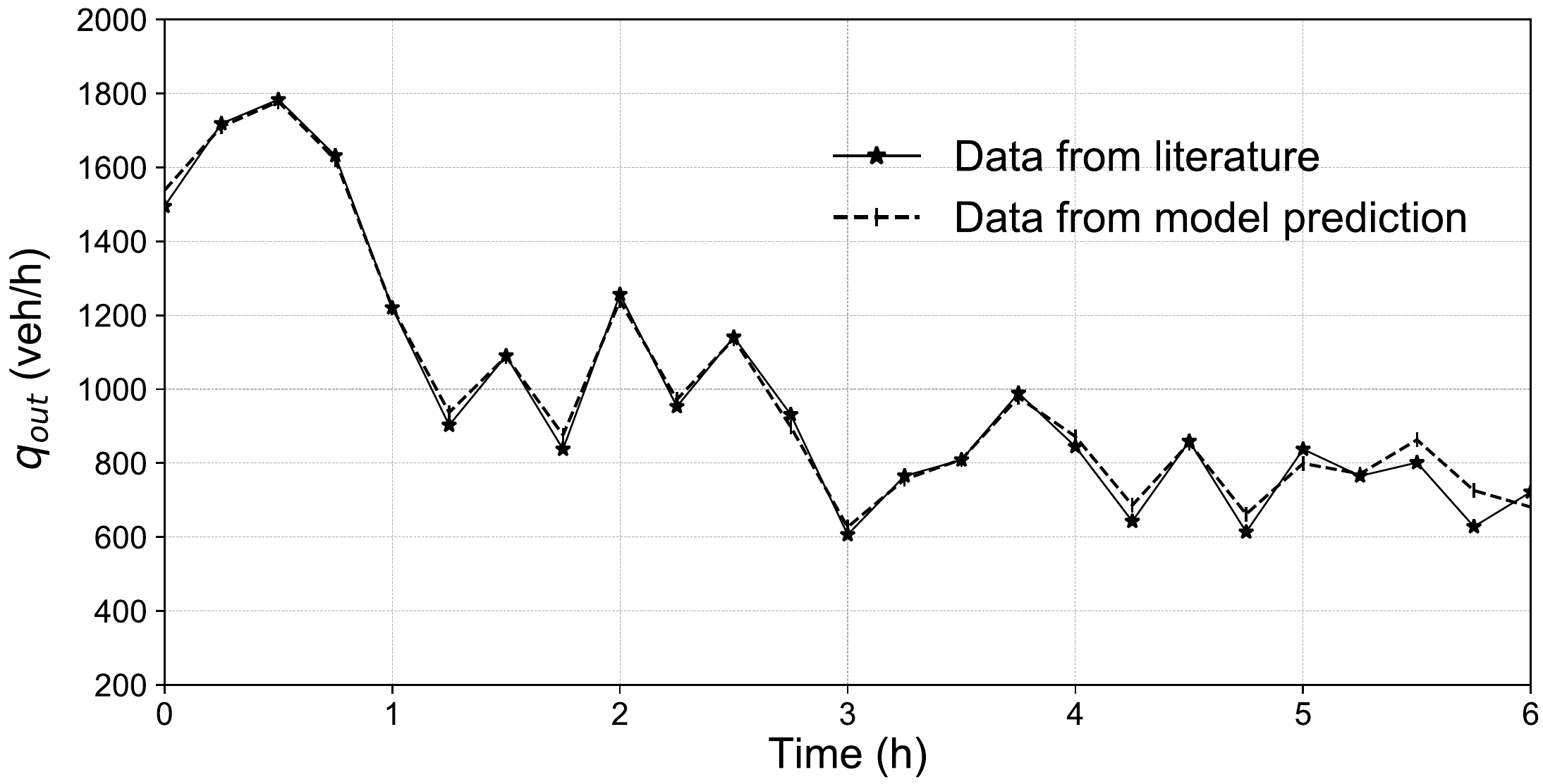}
\caption{Comparison of traffic outflow from model prediction and literature.}
\label{fig:fig11}
\end{figure}
\endgroup

\subsection{Block Layer Model} \label{section:BLM}
The block layer model reflects the interaction of the different system agents in a block, as mathematically illustrated in equations (\ref{eq:eqr1})-(\ref{eq:eqr3}). 
The interdependent variables, shown in bold, exemplify how the outputs of one system agent are inputs of the others.
They are calculated using the system agent models as shown in Section \ref{section:SAM}.
\begin{equation}
\begin{array}{l}
\dot{\vec{S_{E_1}}}(\boldsymbol{LMP^i},\boldsymbol{Sig_E^i}, V^i, I^i, ...)=\\
f_1(\boldsymbol{P_\bold{EV}^i}, \boldsymbol{P_\bold{com}^i},P_\textrm{bui}^i, P_\textrm{PV}^i, P_\textrm{win}^i, ...)
 \label{eq:eqr1}
   \end{array}
\end{equation}
\begin{equation}
\dot{\vec{S_{E_2}}}(\boldsymbol{P_\bold{EV}^i}, \boldsymbol{P_\bold{com}^i})=f_2(\boldsymbol{N_\bold{char}^i}, \boldsymbol{Q_c}, ...)
 \label{eq:eqr6}
\end{equation}
\begin{equation}
\dot{\vec{S_{T_1}}}(\boldsymbol{N_\bold{char}^i})=g_1(\boldsymbol{q_\bold{in,}^i}, \boldsymbol{q_\bold{out}^i}, ...)
 \label{eq:eqr7}
\end{equation}
\begin{equation}
\begin{array}{l}
\dot{\vec{S_{T_2}}}(\boldsymbol{q_\bold{in,}^i}, \boldsymbol{q_\bold{out}^i},U^i, \boldsymbol{t_\bold{w/Com}}, ...)=\\
g_2(C_t^i,U_s^i,\boldsymbol{LMP^i},\boldsymbol{Sig_E^i},\boldsymbol{Sig_T^i},\boldsymbol{Del},...)
 \label{eq:eqr2}
    \end{array}
 \end{equation}
\begin{equation}
\begin{array}{l}
\dot{\vec{S_C}}(\boldsymbol{Q_c}, \boldsymbol{Del}, ...)= \\
h(\boldsymbol{Sig_E^i}, \boldsymbol{Sig_T^i},\boldsymbol{q_\bold{in,}^i}, \boldsymbol{q_\bold{out}^i}, C_c,...)
 \label{eq:eqr3}
     \end{array}
\end{equation} 
     
Operator $\dot{\vec{S}}$ indicates the state variables in the energy, transportation, and communication systems, which are denoted by subscripts $E$, $T$, and $C$, respectively.
Index $i$ designates the node.
In the energy system, the state output variables include node voltage $V$ and line current $I$, as well as the locational marginal price $LMP$ and the energy-related control signals $Sig_E$. These control signals include the buildings, EVs, storage, and renewable energy generations. 
As seen in equation (\ref{eq:eqr1}), the functional inputs to $\dot{\vec{S_{E_1}}}$ include interrelated energy factors $P_\textrm{EV}$ and $P_\textrm{com}$, as well as other load and renewable generation quantities -- such as the building power consumption $P_\textrm{bui}$, $P_\textrm{PV}$ from equation (\ref{eq:eq1}), and $P_\textrm{win}$ from equation (\ref{eq:eq5}).
Further, as shown in equation (\ref{eq:eqr6}), the EV charging active power and communication tower power are determined by the number of charging EVs $N_\textrm{char}$ in equation (\ref{eq:eqr8}) and the communication throughput $Q_c$ in equation (\ref{eq:eq6}), as mentioned in Section \ref{energysystem}.c.

As seen in equation (\ref{eq:eqr7}), $N_\textrm{char}$  correlates to traffic inflow $q_\textrm{in}$ and traffic outflow $q_\textrm{out}$, as exemplified in previous equations (\ref{eq:eq9}). 
In the transportation system shown in equation (\ref{eq:eqr2}), traffic inflow $q_\textrm{in}$ and traffic outflow $q_\textrm{out}$, along with average velocity $U_\textrm{ave}$, road velocity $U$, travel time $t_\textrm{w/Com}$, are determined by the transportation parameters such as road capacity $C_t$, design road velocity $U_s$; energy factors such as the locational marginal price $LMP$ and electrical control signals $Sig_E$; as well as communication factors such as throughput $Q_c$ and time delay $Del$, as exemplified in previous equations (\ref{eq:eq7})-(\ref{eq:eq10}). 
For reference, the empirical traffic flow model included in this paper approximates $U$ as the average road velocity $U_\textrm{ave}$, as shown in equation (\ref{eq:eq7}).
Further, when the communication system is considered, the transportation-related control signals $Sig_T$ affect the travel time $t_\textrm{w/Com}$, which will be elaborated further in Section \ref{section:CLM}. 
In the communication system shown in (\ref{eq:eqr3}), the throughput $Q_c$ and time delay $Del$ affect both the energy and transportation systems, while the transmission threshold $C_c$ affects the communication state as prescribed in (\ref{eq:eq11}).

\begingroup
\setlength{\xfigwd}{\textwidth}
\begin{figure}[h]
\centering
\includegraphics[width=3.5in]{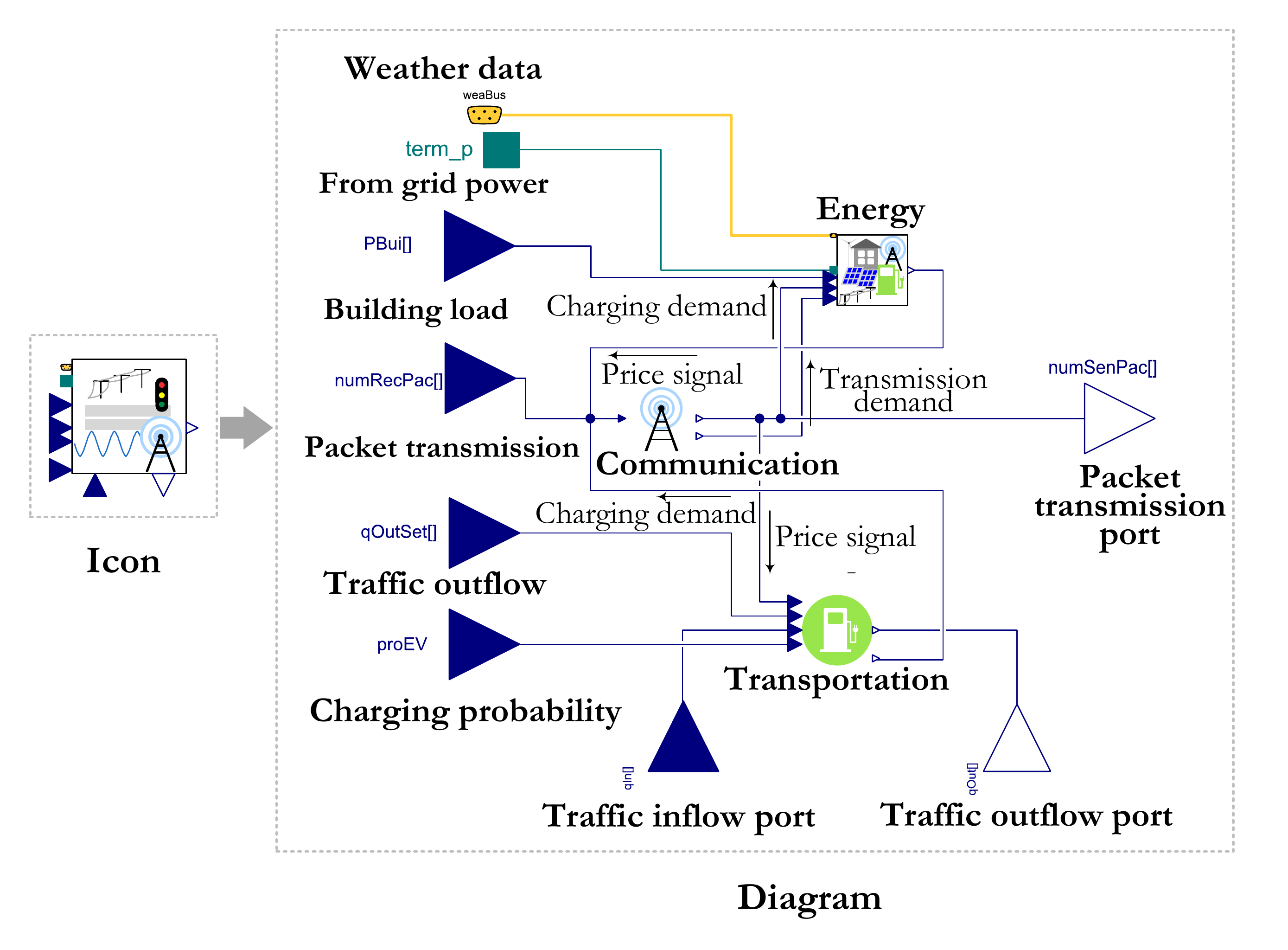}
\caption{Diagram of Modelica model for one block.}
\label{fig:fig12}
\end{figure}
\endgroup

The Modelica implementation  of the block layer model is depicted in Figure \ref{fig:fig12}. As aligned with the block layer in the proposed 3M approach, the transportation agent sends the charging demand signal to the energy system agent via the communication agent. 
Likewise, the energy agent sends back the price/control signal via the communication agent.

Since the energy system supplies power to the communication system, the communication agent sends the throughput demand signal to the energy system agent so that the corresponding energy demand for the communication agent can be calculated. 
In this case study, the communication system in the block is also responsible for transmitting the packets for traffic routing with the interconnected roads through the packet transmission port.

\subsection{Community Layer Model} \label{section:CLM}
The community layer model connects different numbers of blocks. 
Aligned with the 3M approach, different blocks are connected with the power lines and the communicative roads.
The exemplified model shown in Figure \ref{fig:fig13} consists of two blocks. 
The interdependencies on this layer are reflected by these connections of different blocks.
For example, in terms of the energy connection, the voltage between the terminals at Block A and Block B is co-related by linking the terminal connectors with the power lines, which implicitly contains the following mathematical formulation:
\begin{equation}
V^p_{\phi,\textrm{Block A}}-V^n_{\phi,\textrm{Block B}}=Z_\textrm{tot} \cdot I^p_{\phi},
 \label{eq:eqr4}
     \end{equation}
where $V$, $I$, and $Z_\textrm{tot}$ denote the voltage, current, and total impedance, respectively, between the terminals of Block A and Block B at phase $\phi$. 
These correlate to equation (\ref{eq:eqr1}).
At the terminals, $p$ and $n$ represent the positive and negative connectors, respectively.

\begingroup
\setlength{\xfigwd}{\textwidth}
\begin{figure}[h]
\centering
\includegraphics[width=3.5in]{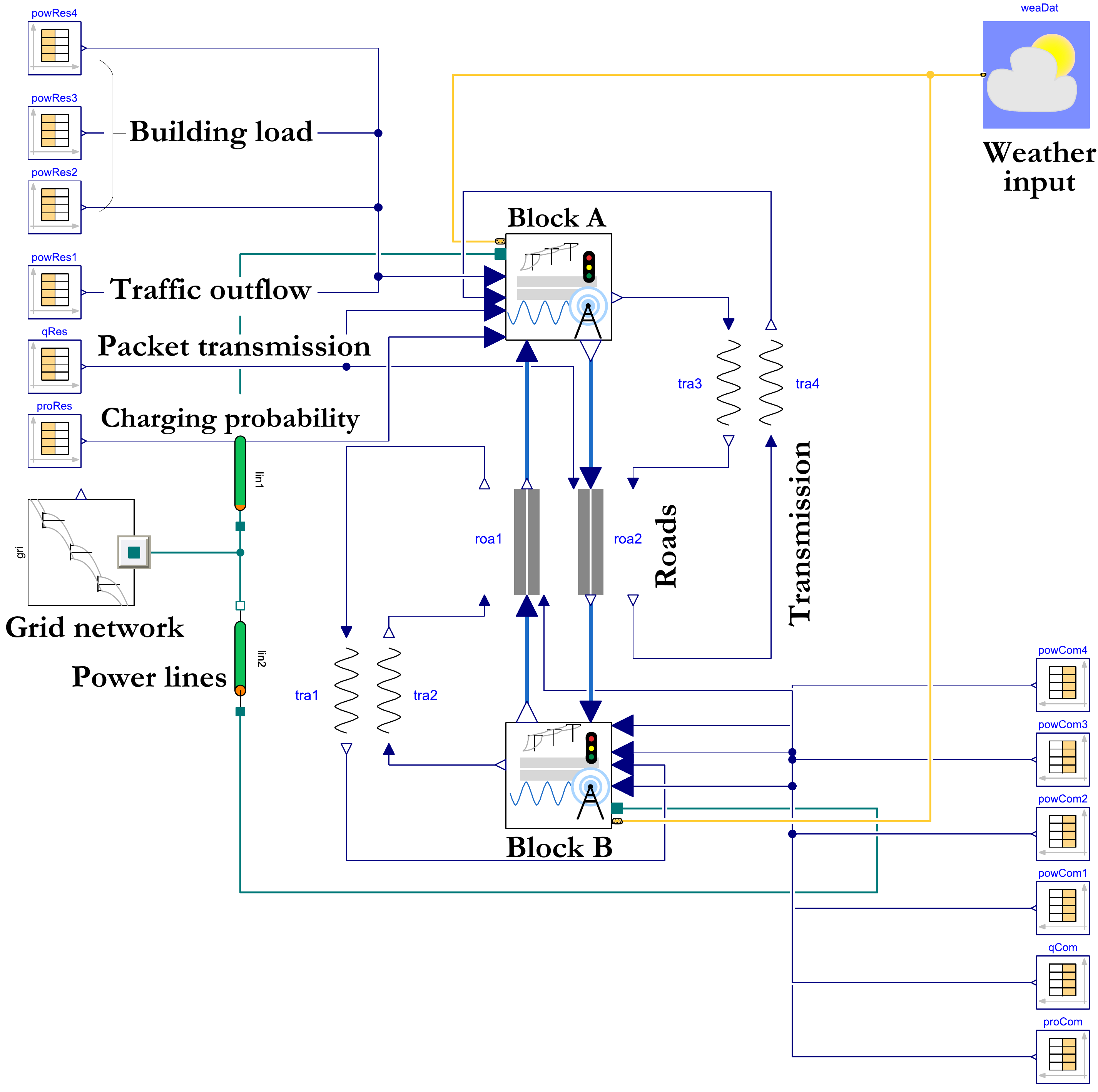}
\caption{Diagram of Modelica model for a community with two connected blocks.}
\label{fig:fig13}
\end{figure}
\endgroup

\begingroup
\setlength{\xfigwd}{\textwidth}
\begin{figure*}[hb!]
\centering
\includegraphics[width=7.2in]{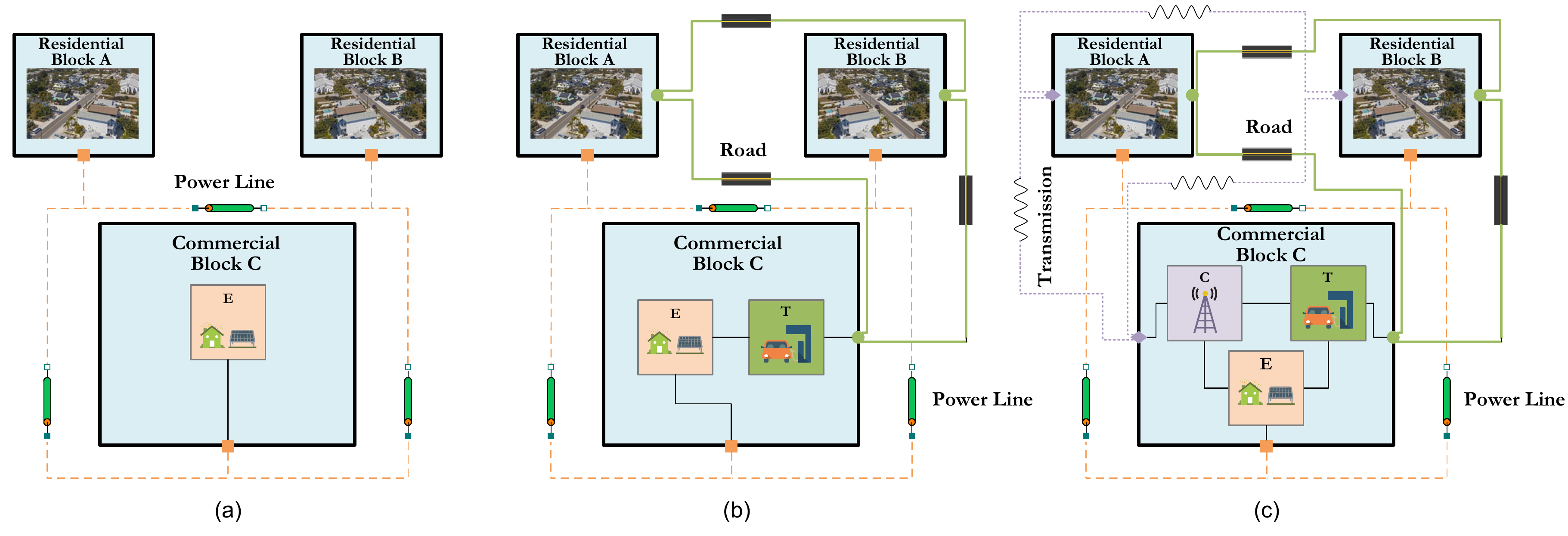}
\caption{Schematics for the three cases: (a) Energy system, (b) Energy + Transportation systems, and (c) Energy + Transportation + Communication systems.}
\label{fig:fig14}
\end{figure*}
\endgroup

\begin{table*} [hb!]
\caption{Detailed model settings in the three cases.}
\label{table2}
\makebox[\linewidth]{
\begin{tabular} {|c|c|c|c|}
\hline
{} & Residential Block 1 & Residential Block 2  & Commercial Block\\
\hline
Weather profile & \multicolumn{3}{c|}{San Francisco, USA}  \\
\hline
PV area ($m^2$) & 20,000 & 30,000  & 50,000\\
\hline
Nominal wind turbine power ($MW$) & \multicolumn{3}{c|}{1}  \\
\hline
Battery capacity ($kWh$) & 4,000 & 5,000  & 6,000\\
\hline
Distribution system type & \multicolumn{3}{c|}{IEEE 16 test feeder}  \\
\hline
Initial EV number & 800 & 800  & 200\\
\hline
Building type & \makecell{600 Residential\\ houses} & \makecell{700 Residential\\ houses}  & \makecell{5 Large offices, 5 strip\\malls, 5 restaurants}\\
\hline
Road Capacity  $(C, U_s, \alpha_1, \alpha_2, \alpha_3)$ & \multicolumn{3}{c|}{\makecell{
$Road_{1,2}$: (350, 30, 1, 1.88, 4.85)\\
$Road_{3,4}$: (1100, 60, 1, 1.88, 7)\\
$Road_{5,6}$: (800,56, 1.4, 1.88, 6.97)
}
}  \\
\hline
Communication system coefficients $(\gamma, C_c)$ & \multicolumn{3}{c|}{\makecell{
$Road_{1,2}$: (0.03,80)\\
$Road_{3,4}$: (0.02,300)\\
$Road_{5,6}$: (0.035,350)
}
}  \\
\hline
\end{tabular}
}
\end{table*}

The communicative roads will exchange routing data with the communication centers in the block. 
Therefore, the performance of the packet exchanges will have an impact on the traffic conditions on the linked roads.
Generically, we represent this interrelation as $Sig_T$ in equations (\ref{eq:eqr2}) and (\ref{eq:eqr3}) above.
While there are many different models to represent this correlating signal, here we adapt the packet loss rate $\gamma$ to represent this signal.
In the current implementation, we assume that the traffic delay time will be proportional to $\gamma$ as follows:
\begin{equation}
t_{\textrm{w/Com}}=(1+\gamma)t_{\textrm{w/oCom}}
 \label{eq:eq12}\end{equation}
where $t_{\textrm{w/Com}}$ and $t_{\textrm{w/oCom}}$ are the travel times with and without consideration of communication, respectively.

\section{Case Studies} \label{section:CasStu}
In this section, three cases are presented in which complexity is gradually added to evaluate the interdependency of different systems and the necessity of integrated modeling. 
Section \ref{section:CasDesPerMea} introduces the basic information of the cases and the performance indicators used to evaluate the multidisciplinary infrastructure models. 
Section \ref{section:cas1} presents the case that only considers the energy system, while the transportation and communication parameters are fixed.
The case study in Section \ref{section:cas2} considers both the energy and transportation system, which intends to investigate the logical interdependencies between the variable charging demand on the energy system and the traffic conditions. 
Lastly, the case study in Section \ref{section:cas3} takes the three systems into account, considering the cyber interdependencies linking the two physical systems. 
These examples are intended to evaluate the intertwined operational performance of energy, transportation, and communication infrastructure in connected communities where large-scale renewable energy generation and full-electrified transportation are adopted.
It is noted that these preliminary models do not fully encapsulate the interdependencies present among the infrastructure systems. In the current case, the charging price is considered to be constant and the charging station control is not taken into account.
As mentioned before, the objective through these case studies it to evaluate the proposed modeling framework for simulating interconnected infrastructure systems for SCC operation.

\subsection{Case Description and Performance Measures}
\label{section:CasDesPerMea}
Figure \ref{fig:fig14} shows the schematics of the three cases. 
The modeled community is composed of two residential blocks and a commercial block. 
As the names imply, the residential blocks mainly consist of residential buildings, while the commercial block is composed of commercial buildings, such as offices, restaurants, and schools. 
Each block has its own renewable generation farm, energy storage (battery), EV charging stations, and communication towers. 
Six one-way roads of the same road type link different blocks. 
The design capacity and design velocity vary between roads. 
The communication system in these cases is dedicated to the traffic routing of the vehicles on the road.
The communication tower in each block exchanges traffic information with the vehicles on the road. 
Thus, the performance of the communication system has a direct effect on the transportation system.
Detailed model settings are shown in Table \ref{table2} for the three cases. 

Several indicators are selected from prior works \cite{b23,b37,b57} to assess the infrastructure performance in connected communities. 
The performance of energy infrastructure includes power load cover factor ($LCF$), peak-valley load ratio ($PVLR$), power system line safety, and power system voltage security. 
Power $LCF$ evaluates the percentage of demand that can be supplied by the renewable energy generation.
Power $PVLR$ is one of the load balancing performance indicators in the power infrastructure, represented by: 
\begin{equation}
PVLR=\frac{P_{\textrm{max}}-P_{\textrm{min}}}{P_{\textrm{max}}}
 \label{eq:eq13}\end{equation}
where $P_{\textrm{max}}$ and $P_{\textrm{min}}$ are the peak and valley power from the grid, respectively.

Power voltage security indicator, $SI_B$, denotes a bus voltage limit placed by the IEEE Standard 519 \cite{b57}: 
\begin{equation}
\begin{split}
 SI_B &= \frac{1}{N_b K} \sum_{i}^{N_b} {\sum_{\kappa}^{K} {g_i(\kappa)}}, \text{ where}\\
g(i)&= \begin{cases}
    v_{\textrm{dev},i}(\kappa)-1.05 &\text{if $v_{\textrm{dev},i}(\kappa)>1.05$}\\    
   0.95-v_{\textrm{dev},i}(\kappa) &\text{if $v_{\textrm{dev},i}(\kappa)<0.95$}\\
   0 &\text{otherwise}
  \end{cases}
\end{split}
 \label{eq:eq14}\end{equation}
where $v_{\textrm{dev},i}(\kappa)$ is the voltage deviation rate of a given line $i$ and a given event $\kappa$ over all events $K$ given a set of $N_b$ buses.

Power line safety indicator, $SI_L$, represents a physical rating on the amount of transferred active power:
\begin{equation}
\begin{split}
 SI_L &= \frac{1}{N_l K} \sum_{i}^{N_l} {\sum_{\kappa}^{K} {f_i(\kappa)}}, \text{ where}\\
f(i)&= \begin{cases}
    P_i(\kappa)-{P_i}^{*} &\text{if $P_i(\kappa)>{P_i}^{*}$}\\
   0 &\text{otherwise}
  \end{cases}
\end{split}
 \label{eq:eq15}\end{equation}
where ${P_i}^{*}$ is the power limit of a given line $i$ and a given event $\kappa$ over all events $K$ given a set of $N_l$ lines.

The performance of transportation and communication systems can be quantified by road congestion and transmission congestion, respectively. 
Road congestion can be gauged by the travel time and traffic flow on the road, while transmission congestion can be reflected by the packet loss or arrival rate. 
The detailed calculation can be seen in Sections \ref{transport} and \ref{communi}.

\subsection{Case 1: Energy System}
\label{section:cas1}
The implementation of Case 1 in Modelica is shown in Figure \ref{fig:fig15}. 
The building loads are hourly load profiles computed using the DOE Commercial Reference Building Models \cite{b58} and Building America House Simulation Protocols \cite{b59}. 
The inputs for the transportation and communication systems are fixed for this case. 
The EV charging profile in one block is prescribed according to the travel demand analysis, and the travel time between the blocks is neglected in the simulation. 

\begingroup
\setlength{\xfigwd}{\textwidth}
\begin{figure}[ht]
\centering
\includegraphics[width=3.2in]{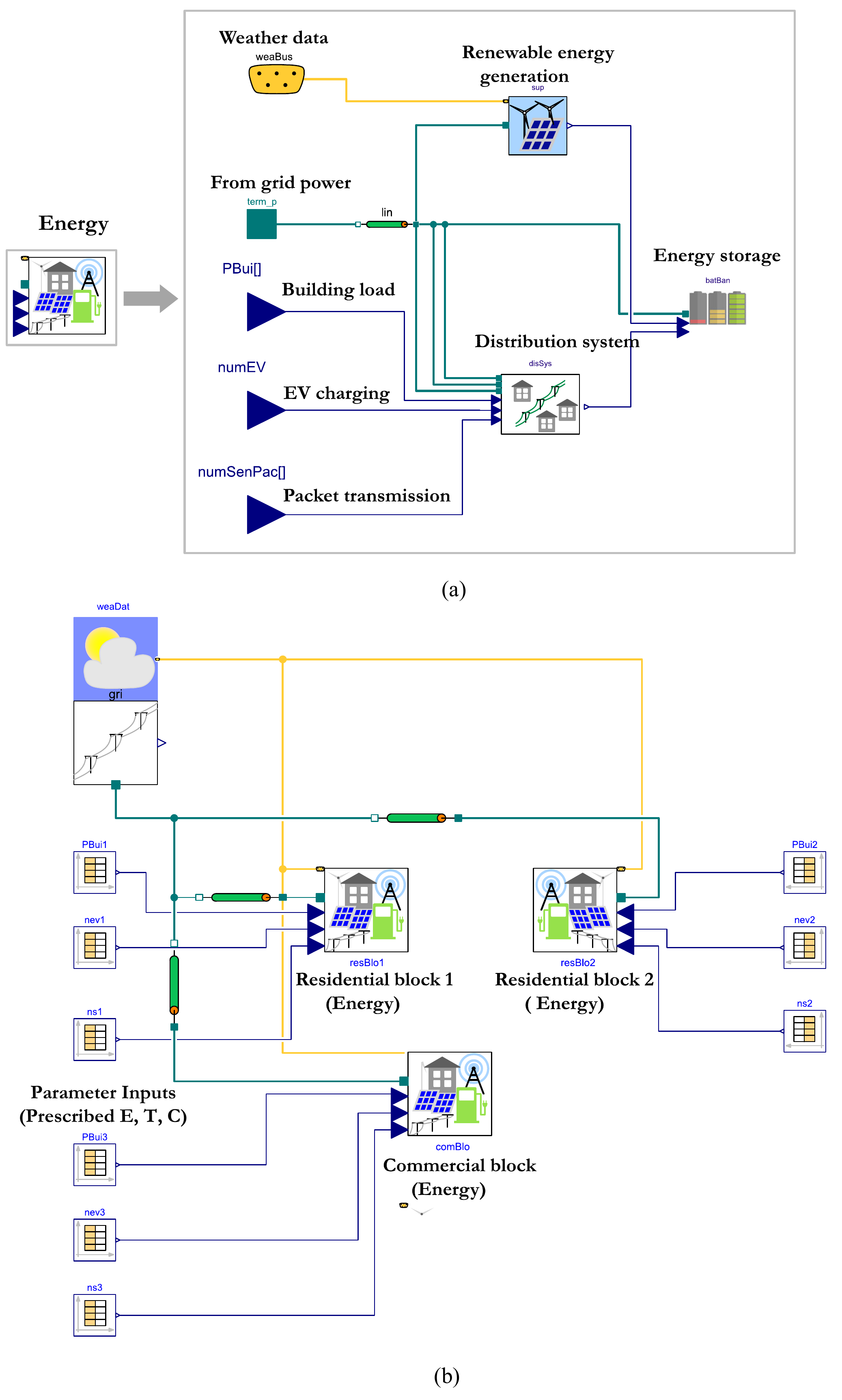}
\caption{Implementation of Case 1: The (a) block icon and (b) community layer diagram for the energy system in Modelica.}
\label{fig:fig15}
\end{figure}
\endgroup

\begingroup
\setlength{\xfigwd}{\textwidth}
\begin{figure}[ht!]
\centering
\includegraphics[width=3.45in]{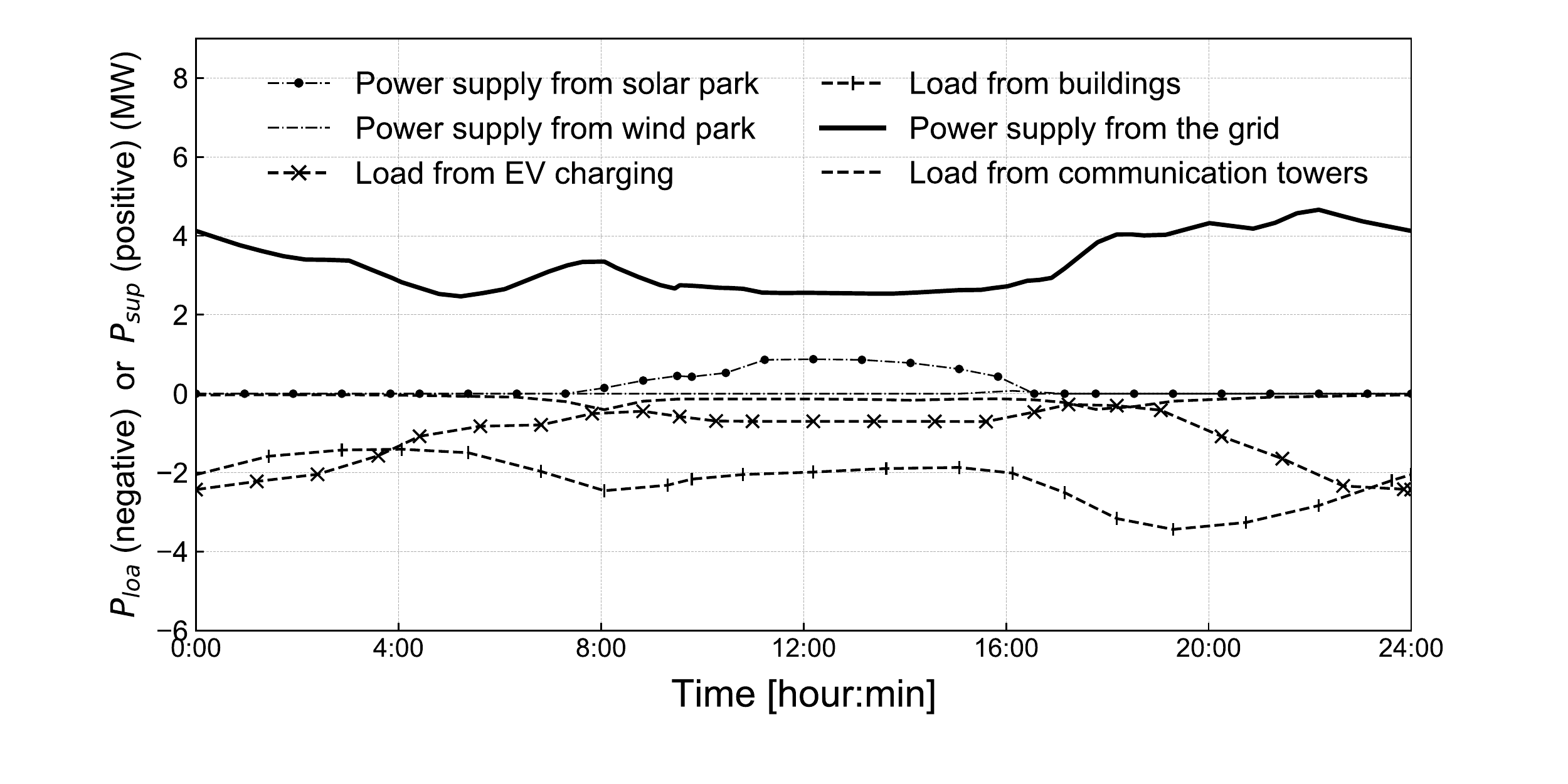}
\caption{Energy system supply and demand profiles for one day in the connected community.}
\label{fig:fig16}
\end{figure}
\endgroup

\newpage
The energy-system simulation results for January 1st in San Francisco are shown in Figure \ref{fig:fig16}. 
From the supply side, the power generated from the PV peaks at noon. 
However, the wind power contributes little during the considered time period. 
From the demand side, the load from the EV charging has a higher demand in the night than during the daytime due to the charging pattern of the residential blocks. 
The load from the building blocks is quite stable compared to the EV charging load pattern. 
It is noted that the load from the communication towers is negligible compared to the load from EV charging and the buildings.

Regarding the power draw from the grid, the peak-valley ratio amounts to 48.79\%. 
This relatively high value is caused by the high EV charging demand at night, which peaks at 22:20, and renewable energy generation during the daytime, which peaks at 12:10. 
However, the duck curve is compromised by the adoption of the battery and the modest power load cover factor (6.92\%). 
The overall results of the energy system indicators are listed in Table \ref{table3}. 

\begin{table} [h]
\caption{Results of the energy system indicators.}
\label{table3}
\centering
\renewcommand{\arraystretch}{1.5}
\begin{tabular}{|c|c|c|c|}
\hline
\multicolumn{4}{|c|}{Energy system indicators}\\
\hline
$PVLR (\%)$ & $LCF (\%)$ & $SI_{B,\textrm{resBlo1}}$ & $SI_{L,\textrm{resBlo1}}$\\
\hline
48.79 & 6.95 & 0.00766  & 0.0097 \\
\hline
\end{tabular}
\end{table}

The $SI_{B,\textrm{resBlo1}}$ and $SI_{L,\textrm{resBlo1}}$ are indices for the power infrastructure performance in terms of the bus safety security and power line safety, respectively. 
Figure \ref{fig:fig17} shows the line voltage deviation rate of three feeders at the distribution system described in Figure \ref{fig:fig6}. 
These results all demonstrate that the power infrastructure has the risk of being overloaded at certain discrete time steps. 
These results suggest that advanced control of coordinating EV charging and building demand response should be taken to avoid the risk.

\begingroup
\setlength{\xfigwd}{\textwidth}
\begin{figure}[hb]
\centering
\includegraphics[width=3.45in]{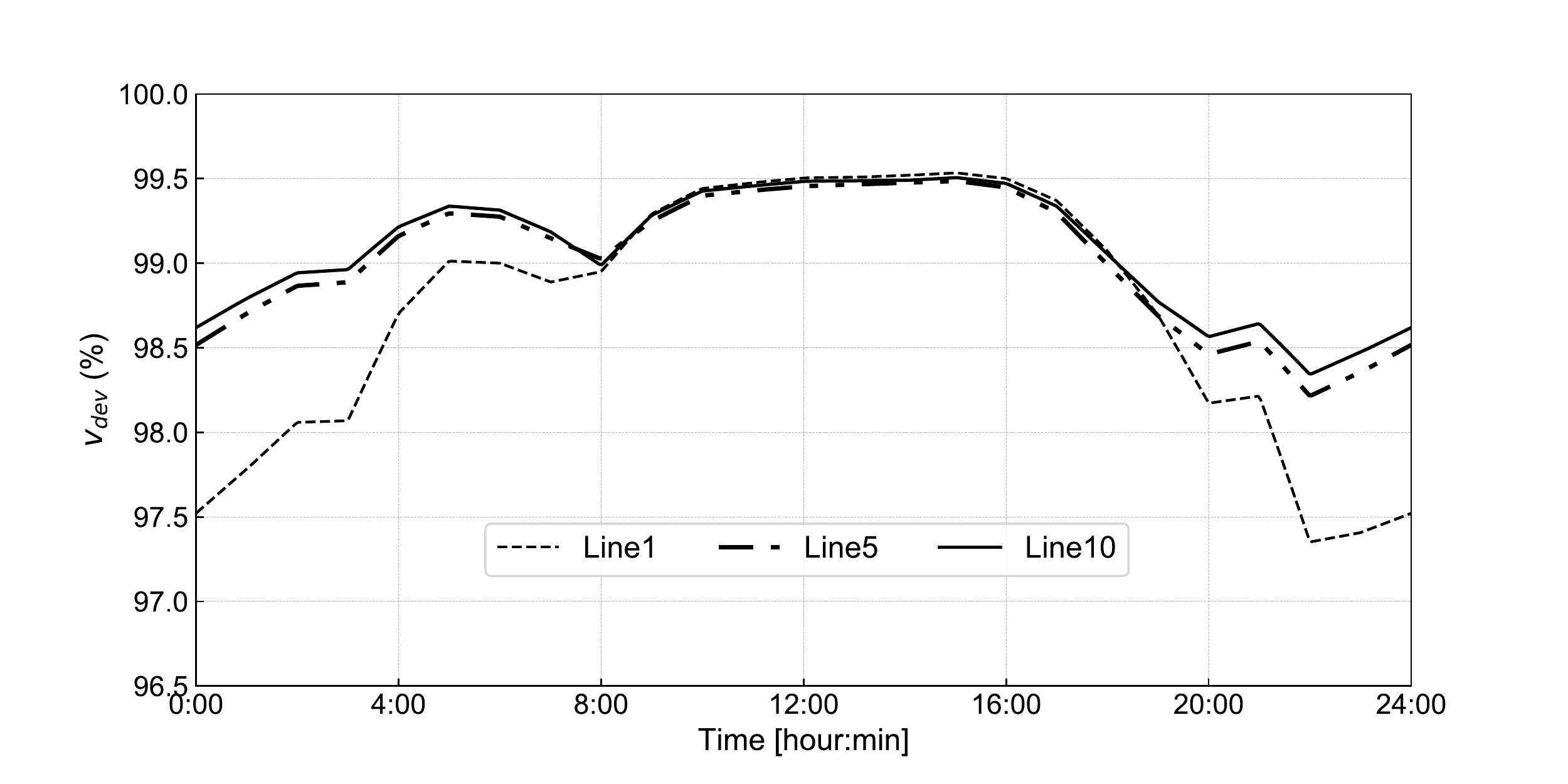}
\caption{Line voltage deviation rate of three feeders at the distribution system described in Figure
\ref{fig:fig6}.}
\label{fig:fig17}
\end{figure}
\endgroup

\subsection{Case 2: Energy + Transportation (E+T) System}

The implementation of Case 2 in Modelica is shown in Figure \ref{fig:fig18}. 
The parameter inputs for the energy system and the communication system settings are the same as Case 1.
However, instead of specifying the EV charging profile, the traffic dynamic is considered in this case based on the traffic outflow profile of the block. 
The traffic outflow profile is determined from the data in the National Cooperative Highway Research Program (NCHRP187) \cite{b60}. 
The quantitative results for the transportation system are illustrated in Figure \ref{fig:fig19}, which shows the average traffic flow profile and the average velocity on different roads. 
The results demonstrate traffic peaks on Road 3 and Road 6 that link the residential blocks and commercial blocks. 

\label{section:cas2}
\begingroup
\setlength{\xfigwd}{\textwidth}
\begin{figure}[ht!]
\centering
\includegraphics[width=3in]{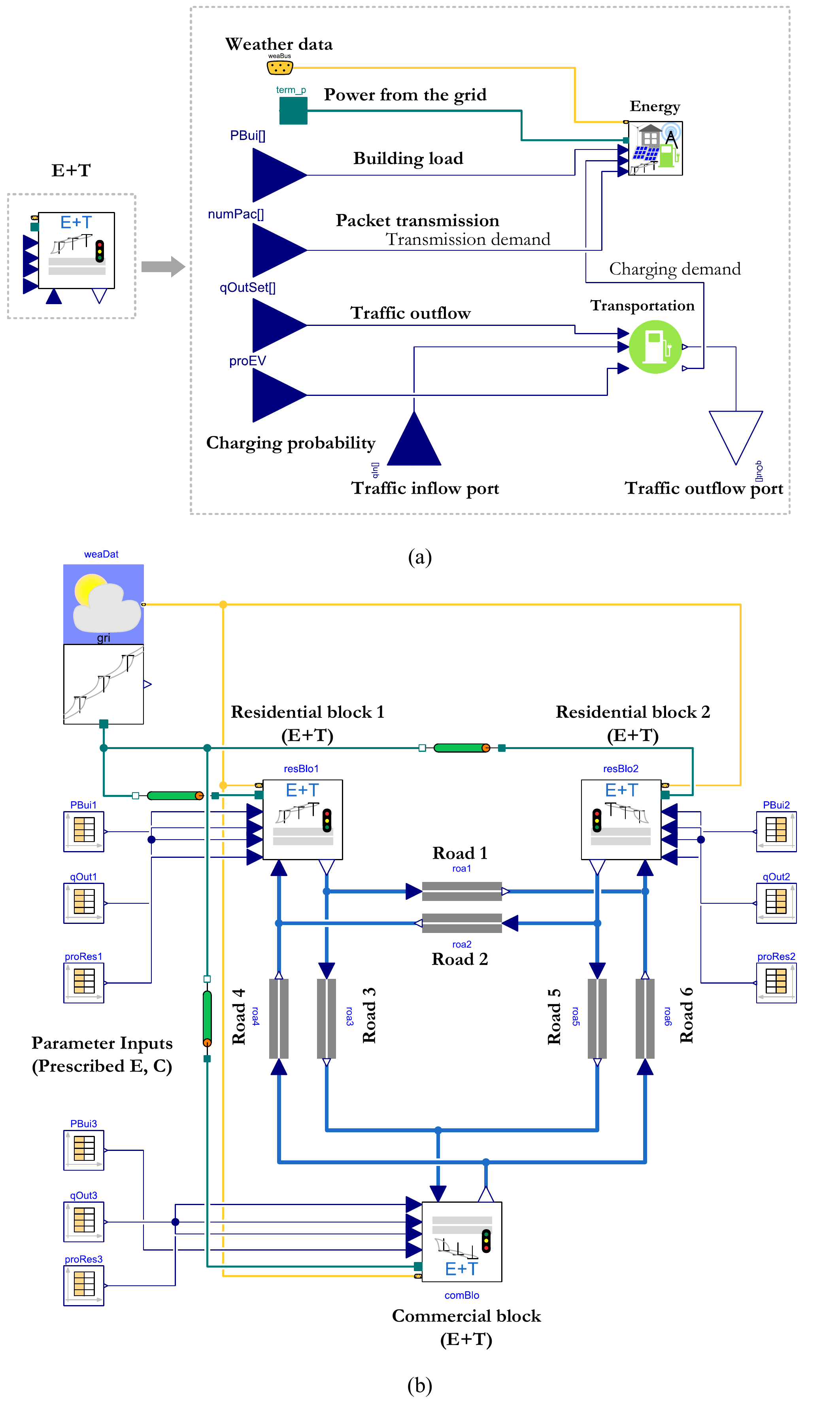}
\caption{Implementation of Case 2: The (a) block icon and (b) community layer diagram for the energy and transportation (E+T) system in Modelica.}
\label{fig:fig18} 
\end{figure}
\endgroup
\begingroup
\setlength{\xfigwd}{\textwidth}
\begin{figure}[h]
\centering
\includegraphics[width=3.35in]{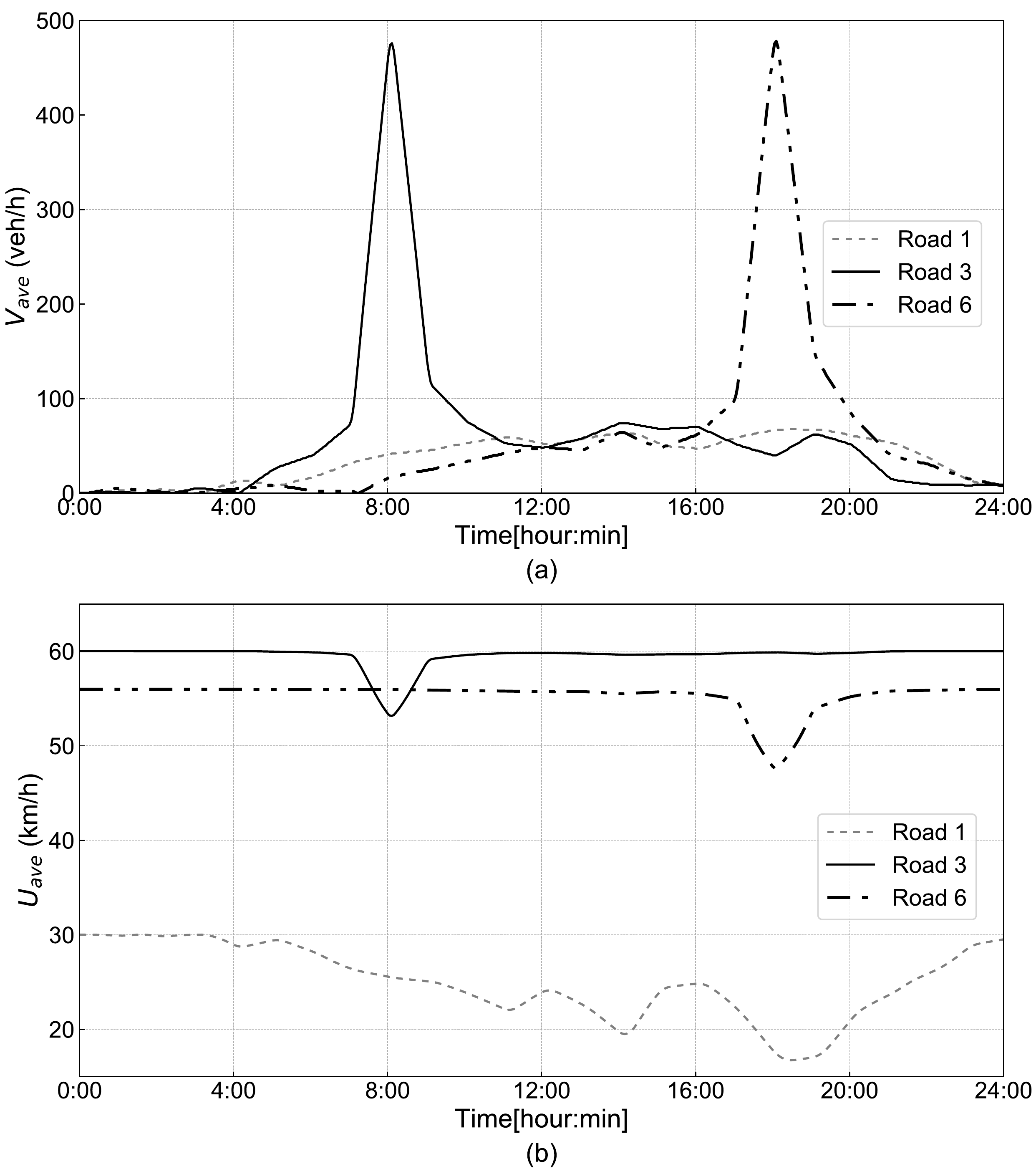}
\caption{Road condition results in the connected community: (a) Average traffic flow profiles and (b) Average velocity profiles on the roads.}
\label{fig:fig19}
\end{figure}
\endgroup

Meanwhile, the average traffic flow on Road 1, which links the two residential blocks, fluctuates less.
At peak commute times (around 8:00 and 18:00), the average traffic flow has a peak value while the average velocity on the road has a minimum value, as can be expected. 
The comparison of the energy results in Case 1 and Case 2 will be analyzed in Section \ref{section:cas2}. 
\newpage

\subsection{Case 3: Energy + Transportation + Communication (E+T+C) System}
\label{section:cas3}
The implementation of Case 3 in Modelica is shown in Figure \ref{fig:fig20}. 
In this case, the model inputs for the energy system and transportation system are the same as Case 2.
Instead of specifying the packet transmission profiles, the packet exchange processes between the road and the corresponding communication tower are modeled as described in Section \ref{transport}. 
As depicted in Figure \ref{fig:fig20}, the transmission happens between the road and the block. 
The assumption is made that the traffic travel time would increase due to the data loss and incomplete transmission of the traffic information. 

\begingroup
\setlength{\xfigwd}{\textwidth}
\begin{figure}[h]
\centering
\includegraphics[width=3.45in]{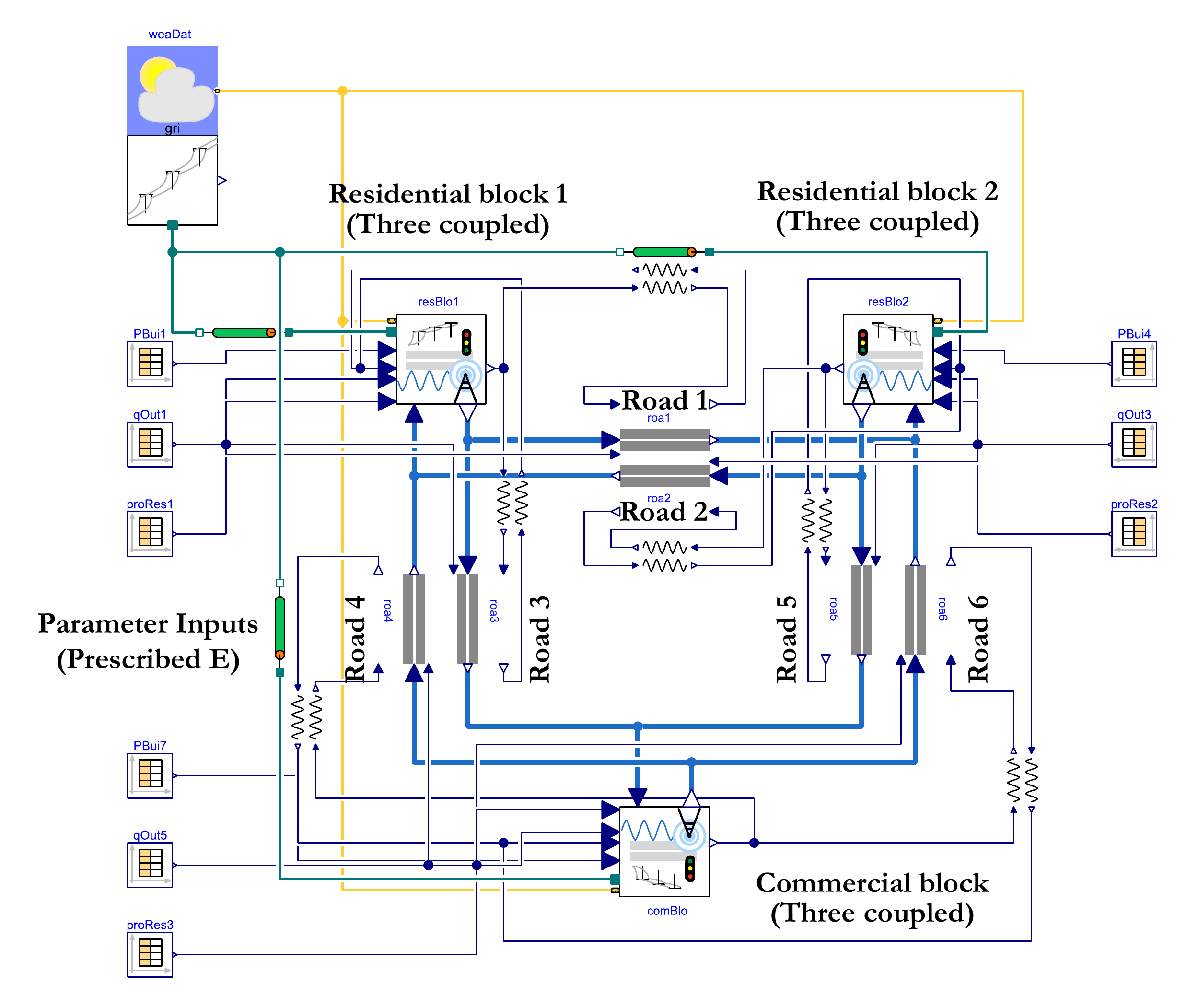}
\caption{Community layer diagram for the E+T+C system in Modelica.}
\label{fig:fig20}
\end{figure}
\endgroup
 
Figure \ref{fig:fig21} shows the interdependencies between the transportation and communication systems. 
Because the communication throughput far exceeds the capacity threshold during commute times, the communication packet loss (shown in Figure \ref{fig:fig21}a) surges when the average traffic flow peaks at the highest commuting times (shown in Figure \ref{fig:fig21}b). 
The packet loss peaks in the morning at Road 3, which is from residential block to commercial block, and in the evening at Road 6, which is from commercial block to residential block.  

\begingroup
\setlength{\xfigwd}{\textwidth}
\begin{figure}[h]
\centering
\includegraphics[width=3.4in]{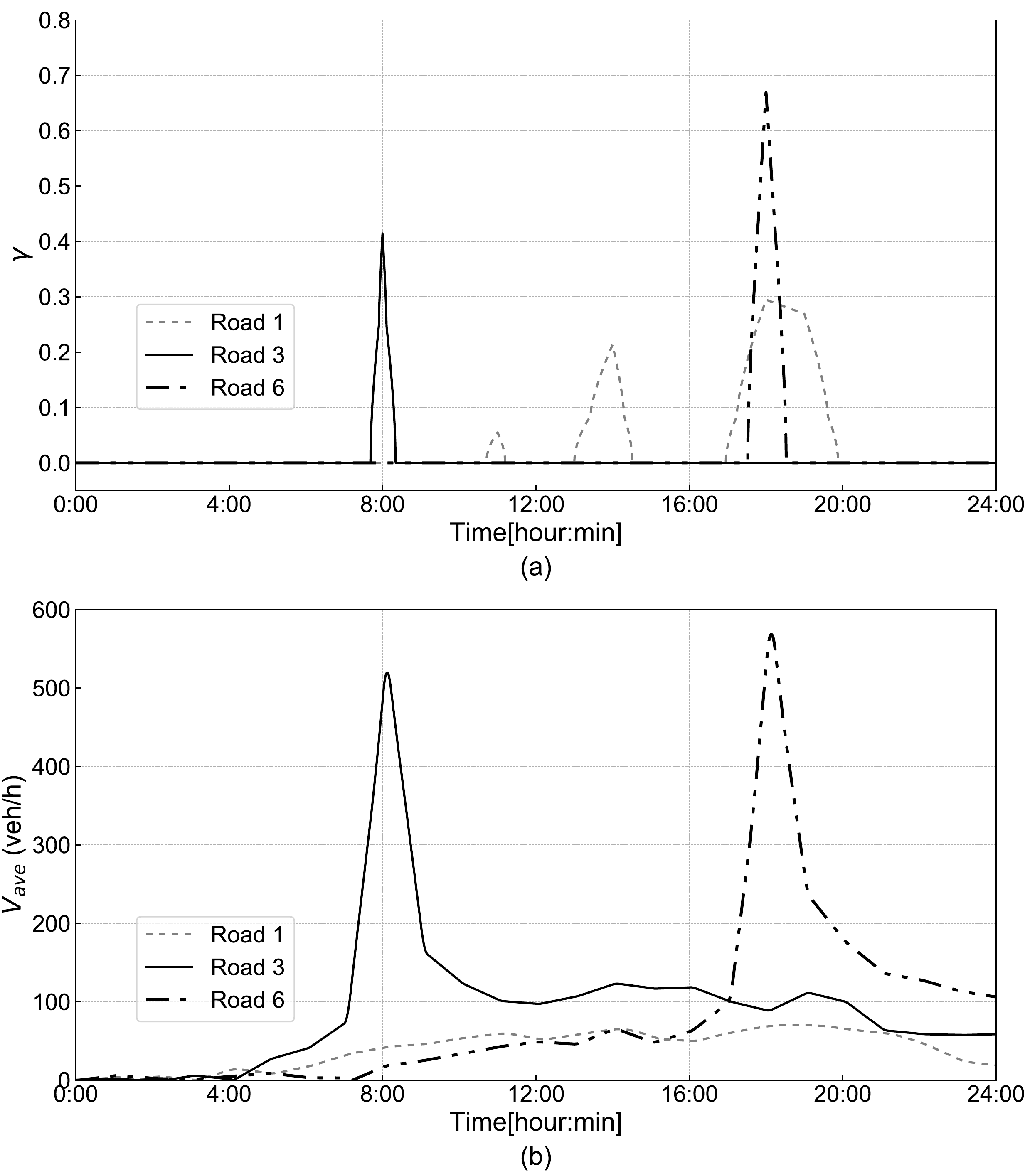}

\caption{ Simulated impact of transportation system on the communication system: (a) packet loss rate for the communication condition and (b) average traffic flow rate for different roads.}
\label{fig:fig21}
\end{figure}
\endgroup

Comparing the simulations from all three cases produces some interesting results. 
Figure \ref{fig:fig22} illustrates the impact of the communication system on the transportation system by comparing the results in Case 2 (E+T) and Case 3 (E+T+C). 
In low traffic hours, the results of both cases almost overlap due to the free flow condition; as such, the communication system has little impact on the traffic condition. 
However, at high traffic hours (around 8:00 and 18:00), the communication system deteriorates the traffic condition due to poor packet arrival rates, as shown in Case 3. 
Compared to Case 2, the average velocities in Case 3 are approximately 10.5\% slower for the morning commute on Road 5 and 3.9\% slower for the evening commute on Road 6. 

\begingroup
\setlength{\xfigwd}{\textwidth}
\begin{figure}[h]
\centering
\includegraphics[width=3.4in]{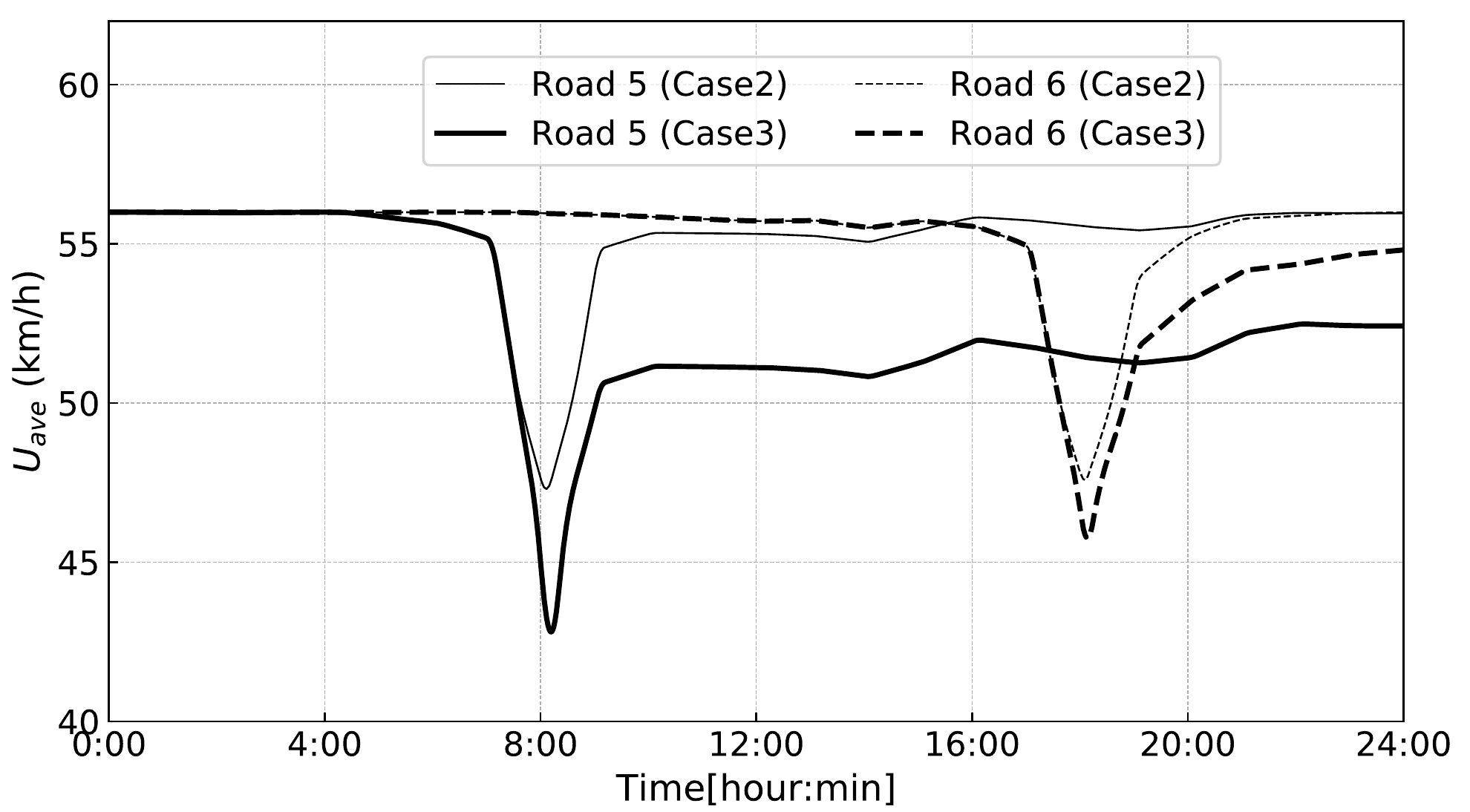}
\caption{Average traffic velocities on different roads with and without modeling of the communication system.}
\label{fig:fig22}
\end{figure}
\endgroup

Beyond the impact of the communication system on transportation, we can analyze the impact of the other two systems on the energy system by comparing the three considered cases. 
Figure \ref{fig:fig23} shows the comparison of the calculated power draw from the grid. 
Clearly, we can see that there is a slight difference on the power demand from the grid during most hours; the average difference is 0.33\% between Case 1 and Case 3. 
The deviation of power draw prediction increases during the peak commuting times (circled). 
This is mainly caused by the collective impacts of the transportation and communication systems. 
The largest deviation ratio of 7\% occurs around 8:00. 
This will have a direct impact on the real-time unit commitment and economic dispatching of the spinning reserves, which help ensure a stable operation of the power grid. 
The power system would result in new location marginal prices. This could in turn change the EV charging pattern and the traffic flow.

\begingroup
\setlength{\xfigwd}{\textwidth}
\begin{figure}[h]
\centering
\includegraphics[width=3.5in]{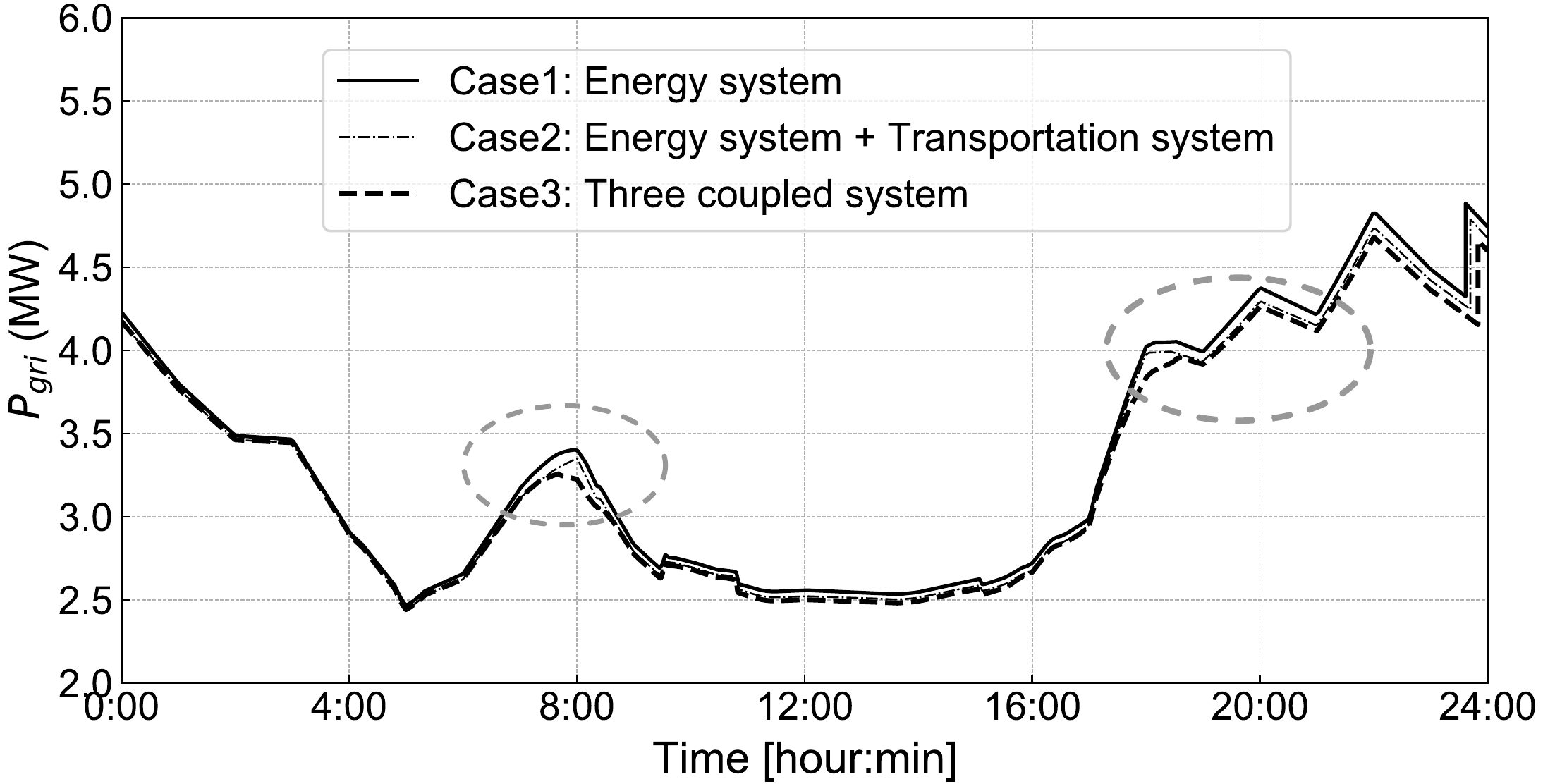}
\caption{Comparison of power draw from the grid in three cases.}
\label{fig:fig23}
\end{figure}
\endgroup

\section{Conclusions} \label{section:Con}
This paper proposes a new flexible and scalable multi-layer, multi-block, multi-agent (3M) approach for modeling several interconnected domains. 
The 3M approach is used to model the interaction of energy, transportation, and communication systems. 
New open source Modelica models are developed accordingly.
The models are applied in three case studies to study the underlying interdependencies of the three systems for evaluating typical SCC operation. 
Simulation results show that the interaction among these three systems should not be neglected. 
In this study, the deviation of the average velocity on the road can be 10.5\% with or without considering the communication system at the peak commute time. 
The deviation of the power draw from the grid can be 7\% with or without considering the transportation and communication systems. 

While these results demonstrate a successful implementation of the proposed 3M modeling framework to study interdependent infrastructure systems, further work is yet needed. 
The case studies included here are the first steps towards modeling of the SCC; as such, they do not fully encapsulate the interedependencies present among these three systems.
Our current focus was to provide the framework which can be later extended.
In the future, we plan to implement the proposed modeling framework to study applications in SCC operation, including dynamic modeling and optimization, resilience analysis, and integrated decision making.

\section*{Acknowledgment}

This research was supported by the National Science Foundation under Awards No. IIS-1802017 and IIS-1633363. BIGDATA: Collaborative Research: IA: Big Data Analytics for Optimized Planning of Smart, Sustainable, and Connected Communities.
This work also emerged from the IBPSA Project 1, an internationally collaborative project conducted under the umbrella of the International Building Performance Simulation Association (IBPSA). 
Project 1 aims to develop and demonstrate a BIM/GIS and Modelica Framework for building and community energy system design and operation.

\newpage

\begin{IEEEbiography}[{\includegraphics[width=1in,height=1.25in,clip,keepaspectratio]{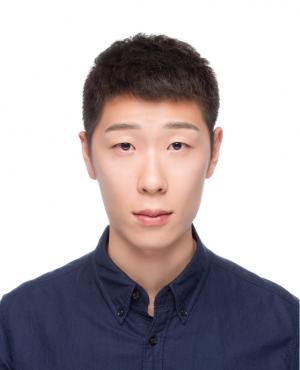}}]{Xing Lu} received a B.S. degree and M.Eng. degree in Power Engineering from Tongji University in 2014 and 2016. He is now a Ph.D. student at the University of Colorado Boulder. His research interests include the modeling and simulation of building systems, urban energy system, and cyber-physical systems.
\end{IEEEbiography}

\begin{IEEEbiography}[{\includegraphics[width=1in,height=1.25in,clip,keepaspectratio]{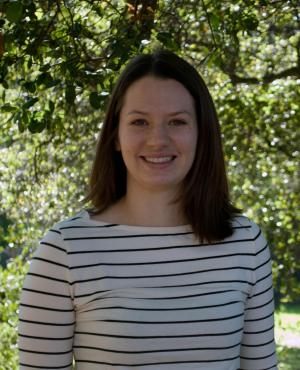}}]{Kathryn Hinkelman} received a B.S. degree in Mechanical Engineering from the University of Denver in 2013. In 2015, she received a M.S. degree in Mechanical Engineering from the University of California at Berkeley. She is currently pursuing a Ph.D. degree at the University of Colorado Boulder. Her research interests include modeling of smart and connected communities, healthy buildings, and sustainable energy systems for low-income populations.
\end{IEEEbiography}

\begin{IEEEbiography}[{\includegraphics[width=1in,height=1.25in,clip,keepaspectratio]{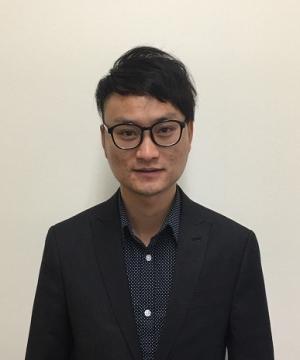}}]{Yangyang Fu} is currently pursuing a Ph.D. degree in architectural engineering in University of Colorado Boulder. His research interests include the application of data mining and machine learning techniques in the building energy area, optimal design and operation of the chiller plant system, discrete-event and continuous system modelling, and numerical solutions of hybrid systems. 
\end{IEEEbiography}

\begin{IEEEbiography}[{\includegraphics[width=1in,height=1.25in,clip,keepaspectratio]{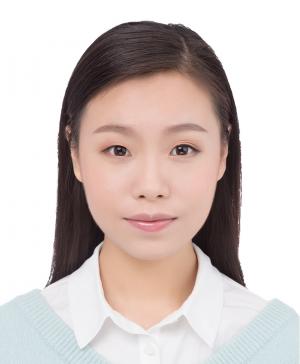}}]{Jing Wang} received a B.S. degree and M.S. degree from Tongji University in 2009 and 2013. She is currently pursuing a Ph.D. degree in architectural engineering. Her current work focuses on the development of a computational framework for resilient coastal cities.
\end{IEEEbiography}

\begin{IEEEbiography}[{\includegraphics[width=1in,height=1.25in,clip,keepaspectratio]{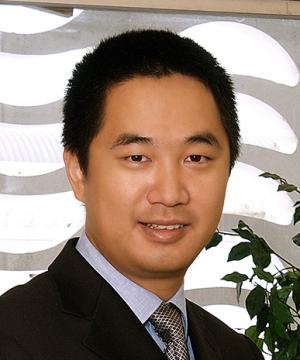}}]{Wangda Zuo} received a B.S. degree and M.Eng. degree from Chongqing University in 2000 and 2003. He also earned a M.Sc. degree from Friedrich-Alexander-University Erlangen-Nuremberg in 2005. He received a Ph.D degree from Purdue University in 2010.  
He is now an Associate Professor at the University of Colorado Boulder. He was the recipient of the IBPSA-USA Emerging Professional Award, Eliahu I. Jury Early Career Research Award, ASHRAE Distinguished Service Award, Provost Research Award, and SEEDS Leadership Award. 
He also served as Treasurer and Affiliate Director (representing USA) of the International Building Performance Simulation Association (IBPSA); Former Chair of IBPSA-USA Research Committee; Chair and Voting Member of ASHRAE Technical Committee (TC) 7.4 Exergy Analysis for Sustainable Buildings; Vice Chair of ASHRAE (TC) 4.10 Indoor Environment Modeling; and Former Voting Member of ASHRAE TC 4.7 Energy Calculations. Prior to joining CU Boulder, he was an Assistant Professor at the University of Miami and Computational Research Scientist at Lawrence Berkeley National Laboratory.
\end{IEEEbiography}

\begin{IEEEbiography}[{\includegraphics[width=1in,height=1.25in,clip,keepaspectratio]{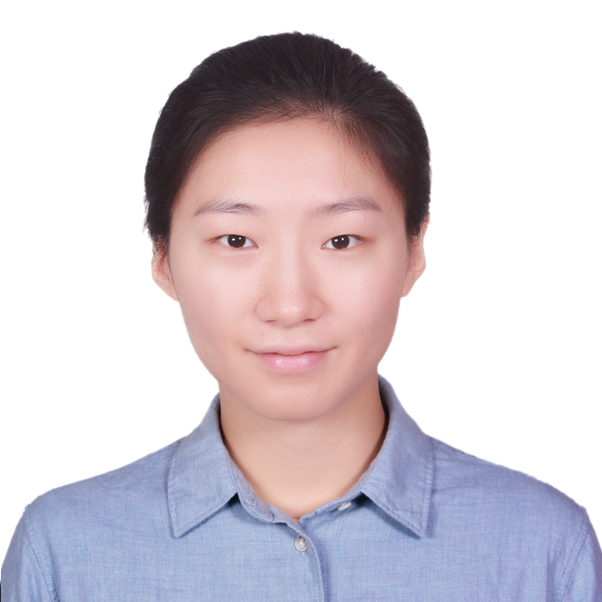}}]{Qianqian Zhang} (S'16) is currently a Ph.D. student in Bradley Department of Electrical \& Computer Engineering at Virginia Tech, Blacksburg, VA, USA. She works in the NetSciWiS research group at Wireless@Virginia Tech, with Dr. Walid Saad as her adviser. She received her B.S. degree in Telecommunication Engineering with Management from Beijing University of Posts and Telecommunications (BUPT), Beijing in 2015. Her research interests include millimeter wave communications, internet of things, game theory, and machine learning.
\end{IEEEbiography}

\begin{IEEEbiography}[{\includegraphics[width=1in,height=1.25in,clip,keepaspectratio]{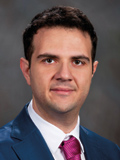}}]{Walid Saad}  (S'07, M'10, SM'15, F'19) 
received his Ph.D degree from the University of Oslo in 2010. He is currently an Associate Professor at the Department of Electrical and Computer Engineering at Virginia Tech, where he leads the Network Science, Wireless, and Security laboratory. His research interests include wireless networks, machine learning, game theory, security, unmanned aerial vehicles, cyber-physical systems, and network science. Dr. Saad is a Fellow of the IEEE and an IEEE Distinguished Lecturer. He is also the recipient of the NSF CAREER award in 2013, the AFOSR summer faculty fellowship in 2014, and the Young Investigator Award from the Office of Naval Research (ONR) in 2015. He was the author/co-author of seven conference best paper awards at WiOpt in 2009, ICIMP in 2010, IEEE WCNC in 2012, IEEE PIMRC in 2015, IEEE SmartGridComm in 2015, EuCNC in 2017, and IEEE GLOBECOM in 2018. He is the recipient of the 2015 Fred W. Ellersick Prize from the IEEE Communications Society, of the 2017 IEEE ComSoc Best Young Professional in Academia award, and of the 2018 IEEE ComSoc Radio Communications Committee Early Achievement Award. From 2015-2017, Dr. Saad was named the Stephen O. Lane Junior Faculty Fellow at Virginia Tech and, in 2017, he was named College of Engineering Faculty Fellow. He currently serves as an editor for the IEEE Transactions on Wireless Communications,  IEEE Transactions on Mobile Computing, IEEE Transactions on Cognitive Communications and Networking, and IEEE Transactions on Information Forensics and Security. He is an Editor-at-Large for the IEEE Transactions on Communications.

\end{IEEEbiography}

\end{document}